\documentclass[journal]{IEEEtran}
\usepackage{float}
\usepackage{graphicx}
\usepackage{subfigure}
\usepackage{epstopdf}
\usepackage{color}
\usepackage{amssymb}
\usepackage{amsmath}
\usepackage{algorithm}
\usepackage{algorithmic}
\usepackage{bm}
\usepackage[colorlinks]{hyperref}

\begin{document}
\title{FAS vs. ARIS: Which Is More Important for FAS-ARIS Communication Systems?}
\author{Junteng Yao, Liaoshi Zhou, Tuo Wu, Ming Jin, Chongwen Huang, and Chau Yuen, \emph{Fellow, IEEE}

\thanks{J. Yao, L. Zhou, and M. Jin are with the faculty of Electrical Engineering and Computer Science, Ningbo University, Ningbo 315211, China (E-mail: $\rm \{yaojunteng, 2311100202, jinming\}@nbu.edu.cn$).}

\thanks{T. Wu and C. Yuen are with the School of Electrical and Electronic Engineering, Nanyang Technological University, 639798, Singapore (E-mail: $\rm \{tuo.wu, chau.yuen\}@ntu.edu.sg$).}

\thanks{C. Huang is with College of Information Science and Electronic Engineering, Zhejiang University, Hangzhou 310027,
China, and Zhejiang Provincial Key Laboratory of Info. Proc., Commun. $\&$ Netw. (IPCAN), Hangzhou 310027, China. (Email: $\rm chongwenhuang@zju.edu.cn$).}
}

\maketitle
\begin{abstract}
In this paper, we investigate the question of which technology, fluid antenna systems (FAS) or active reconfigurable intelligent surfaces (ARIS), plays a more crucial role in FAS-ARIS wireless communication systems. To address this, we develop a comprehensive system model and explore the problem from an optimization perspective. We introduce an alternating optimization (AO) algorithm incorporating majorization-minimization (MM), successive convex approximation (SCA), and sequential rank-one constraint relaxation (SRCR) to tackle the non-convex challenges inherent in these systems.  Specifically, for the transmit beamforming of the BS optimization, we propose a closed-form rank-one solution with low-complexity. For the optimization the positions of fluid antennas (FAs) of the BS, the Taylor expansions and MM algorithm are utilized to construct the effective lower bounds and upper bounds of the objective function and constraints, transforming the non-convex optimization problem into a convex one. Furthermore, we use the SCA and SRCR to optimize the reflection coefficient matrix of the ARIS and effectively solve the rank-one constraint. Simulation results reveal that the relative importance of FAS and ARIS varies depending on the scenario: FAS proves more critical in simpler models with fewer reflecting elements or limited transmission paths, while ARIS becomes more significant in complex scenarios with a higher number of reflecting elements or transmission paths. Ultimately, the integration of both FAS and ARIS creates a win-win scenario, resulting in a more robust and efficient communication system. This study underscores the importance of combining FAS with ARIS, as their complementary use provides the most substantial benefits across different communication environments.
\end{abstract}
\section{Introduction}
\IEEEPARstart{A}{s} the development of sixth-generation (6G) networks accelerates, the limitations of traditional antenna systems from 5G, such as massive multiple input multiple output (MIMO) technology, are becoming evident, particularly in their inability to meet the stringent quality of service (QoS) requirements expected in 6G. In response, fluid antenna systems (FAS) \cite{KKWong22,KKWong2023,JYao24, WMa23,LZhu23,M.Khammassi23,David24,Vega2024} are gaining prominence as a crucial innovation. These systems dynamically adjust antenna positions, greatly enhancing data transmission rates and optimizing spatial resource utilization for next-generation networks.

Moving beyond traditional fixed-position antennas (FPAs), fluid antenna systems (FAS) introduce a revolutionary approach to wireless communications. These systems facilitate the active reconfiguration of the wireless propagation channel, thereby expanding spatial degrees of freedom (DoFs) and significantly enhancing communication rates \cite{LZhu24,XLai24,KKWong23,NWaqar23,KKWong231}. As a result, FAS have become essential components of various advanced wireless applications, including multi-user systems \cite{HQin24}, over-the-air computation (AirComp) systems \cite{DZhang24}, and non-orthogonal multiple access (NOMA) systems \cite{JZheng24}.

However, a significant challenge in FAS-assisted communication systems arises from the need to adjust the fluid antenna (FA) to optimal positions for selecting the best available channels. This adjustment is not only complex but also requires sophisticated channel estimation techniques, contributing to considerable operational overhead \cite{Xu1}. Achieving optimal antenna placement swiftly demands algorithms with high complexity and overhead, complicating the system design further. Beyond these challenges in algorithm development, this situation prompts a reevaluation of solutions at the physical layer. Fortunately, the innovative concept of reconfigurable intelligent surfaces (RISs) \cite{Wu1,Wu2,Wu3,Zhi1,Zhi2}, also known as intelligent reflecting surfaces (IRSs) \cite{Yao1}, emerges as a viable alternative. RIS technology can potentially simplify the channel adaptation process by modifying the signal phase dynamically, thus reducing the dependency on rapid and precise FA positioning for optimal channel selection.

Comprising numerous cost-effective, passive reconfigurable elements, RISs can intelligently adjust the phase shifts of impinging waves under a controller's direction \cite{Zhi1}. These surfaces operate without the need for radio frequency (RF) chains and digital signal processing circuits, supporting a compact design that simplifies installation on building facades and indoor environments. Consequently, RIS technology can significantly alleviate these challenges, aiding in channel control and compensating for communication performance losses due to suboptimal FA positioning.

Inspired by the significant advantages of RIS in FA-assisted communication systems, FAS-RIS communication systems have attracted considerable research interest. Specifically, regarding performance analysis, the authors in \cite{FASRIS1} investigated the performance of RIS-aided FAS systems. Expanding further, the study in \cite{Yao2} proposed a comprehensive framework that includes performance analysis and throughput optimization. From an optimization standpoint, research in \cite{FASRIS2} explored low-complexity beamforming design for RIS-assisted FAS systems, while the work in \cite{FASRIS3} focused on sum-rate optimization for RIS-aided multiuser MAS systems.

Nevertheless, the deployment of RIS can lead to limited performance enhancements due to the ``multiplicative fading" or ``double fading" effect. This occurs because the signals in RIS-aided systems must traverse the cascaded channel, where the path-loss for the cascaded link is calculated as the product of the path losses at each segment, resulting in a total path-loss significantly higher than that of a direct link \cite{ARIS1}. To overcome this, the authors of \cite{ARIS1} and \cite{ARIS2} recently proposed a new  architecture, namely active RIS (ARIS). Unlike passive RISs that merely adjust the phase of reflected signals, ARISs are equipped with active reflective amplifiers that also amplify the signals. This innovation has shown that ARISs  can achieve better gain compared with the passive RISs. Furthermore, these ARISs are designed to be relatively thin, which facilitates easy deployment on building facades and within indoor spaces, similar to passive RISs but without requiring RF chains and digital signal processing circuits. Although ARISs incur higher hardware power consumption due to their amplifiers, they have been shown to outperform passive RISs when the number of reflecting elements is small and the system power budget is large \cite{Zhi3,BWei2024,MHKhoshafa2022,ZShi2023}. Therefore, ARISs not only mitigate the multiplicative fading effect but also retain many benefits of passive systems.

Leveraging the advantages of ARISs over passive RISs, ARISs could substantially enhance the efficacy of FAS-assisted communication systems. By incorporating ARISs, these systems could see notable improvements in signal strength and communication performance, primarily due to the ARISs' capacity to both mitigate multiplicative fading effects and amplify signals. This enhancement is particularly crucial in scenarios where direct signal paths are obstructed, significantly boosting the range and reliability of communications.  However, in different deployment scenarios, it is essential to determine which technology, FAS or ARIS, plays a more critical role, and to what extent each contributes to overall system performance. Despite the promising potential of these technologies, there is a lack of comprehensive research that quantitatively analyzes their relative importance in various conditions. Therefore, it is necessary to compare which is more important from an optimization perspective. Such an analysis is crucial, as it would provide valuable insights for future 6G wireless network managers, guiding decisions on the optimal deployment of FAS and ARIS to maximize communication efficiency.

Optimizing FAS-ARIS communication systems introduces complex challenges that extend beyond merely positioning FAs. The comprehensive optimization process also involves addressing the beamforming optimization of the FAs and the phase adjustments within the ARIS. This multi-faceted optimization issue presents a non-trivial challenge, as it requires simultaneous consideration of various interacting components. Furthermore, the unique properties of ARIS complicate the beamforming optimization, increasing the complexity of developing effective solutions. Traditional algorithms often fail to provide closed-form solutions for beamforming in such integrated systems, further escalating the difficulty of achieving optimal configurations. This complexity underscores the need for innovative optimization strategies tailored to the distinct dynamics of FAS-ARIS systems.

Against this background, this work introduces a FAS-ARIS downlink communication system, featuring a BS equipped with multiple FAs transmitting signals through an ARIS to a user equipment (UE) with a FPA. Within this system, the ARIS plays a crucial role in enhancing system performance. Simultaneously, the integration of FAS further elevates overall communication capabilities by dynamically selecting optimal radiation patterns to maximize the achievable rate at the UE. We detail our contributions as follows:

\begin{itemize}
\item To compare the importance of FAS and ARIS within the system, we investigate a FAS-ARIS downlink communication setup consisting of a BS, a UE, and an ARIS. The BS is equipped with multiple FAs and transmits signals via the ARIS to the UE, which is equipped with a single FPA. Based on this system configuration, we formulate an optimization problem aimed at maximizing the achievable rate at the UE, while adhering to the power constraints of both the ARIS and the BS. This involves jointly optimizing the BS's transmit beamforming, the positioning of the BS's transmit FAs, and the ARIS's reflection coefficient matrix. The complexity of this task is compounded by the highly non-convex nature of the problem, due to the intricate coupling of the optimization variables within both the objective function and the constraints.

\item To tackle this highly non-convex optimization problem, we propose an alternating optimization (AO)-based algorithm that integrates several advanced techniques:  majorization-minimization (MM), successive convex approximation (SCA), semidefinite relaxation (SDR) and sequential rank-one constraint relaxation (SRCR). Specifically, within the ARIS system, our beamforming optimization reveals an optimal closed-form rank-one solution that satisfies all constraints, facilitating a low-complexity implementation. The employed MM algorithm and the Taylor expansions are employed to iteratively optimize the positions of the FAs, enhancing system configuration adaptability. Lastly, the optimization of the reflection coefficient matrix at the ARIS is efficiently handled using the SRCR algorithm, which helps avoid the potential non-convergence issues associated with SDR.

\item Simulation results confirm that deploying FAs in RIS-assisted communication systems significantly improves performance compared to FPAs. Furthermore, the integration of ARIS not only enhances the performance of FA-assisted communication systems but also outperforms passive RIS systems. These findings highlight the complementary nature of FAS and ARIS, where FAS proves particularly important in scenarios with fewer reflecting elements or limited transmission paths, and ARIS becomes more crucial in more complex environments with a higher number of reflecting elements or transmission paths. The combined use of FAS and ARIS creates a win-win scenario, resulting in a more robust and efficient communication system, further validating the efficacy of the integrated FAS-ARIS systems.
 \end{itemize}
\textit{Notations}: The trace, rank, Frobenius norm, 2-norm, conjugate transpose, conjugate, and transpose of the matrix $\mathbf{A}$ are denotes as $\mathrm{Tr}(\mathbf{A})$, $\mathrm{rank}(\mathbf{A})$, $\left\|\mathbf{A}\right\|_F$, $\left\|\mathbf{A}\right\|_2$, $\mathbf{A}^H$, $\mathbf{A}^{\ast}$, and $\mathbf{A}^T$, respectively; $[\mathbf{A}]_{m, n}$ represents the $m$th row and $n$th column element of matrix $\mathbf{A}$; $\mathbf{A}\succeq (\succ) \mathbf{0}$ indicates that the matrix $\mathbf{A}$ is positive semidefinite (positive definite); $\Re\{x\}$ means the real part of $x$; The term $\mathbb{C}^{m\times n}$ denotes a complex matrix with size $m \times n$; $\lambda_{\max}(\mathbf{A})$ denotes the maximum eigenvalue of matrix $\mathbf{A}$. $|a|$ and $\angle a$ denote the amplitude and phase of complex scalar $a$, respectively;  $\mathbb{E}\left\{\cdot\right\}$ denotes the expectation operation; $\mathrm{diag}\{\cdot\}$ denotes the diagonalization operation; $\mathcal{CN}(\mathbf{0}, \mathbf{I})$ denotes the distribution of a circularly symmetric complex Gaussian vector with mean $\mathbf{0}$ and covariance $\mathbf{I}$.

\section{ System Model and Problem Formulation}
\begin{figure}[t]
\centering
\includegraphics[width=3.5 in]{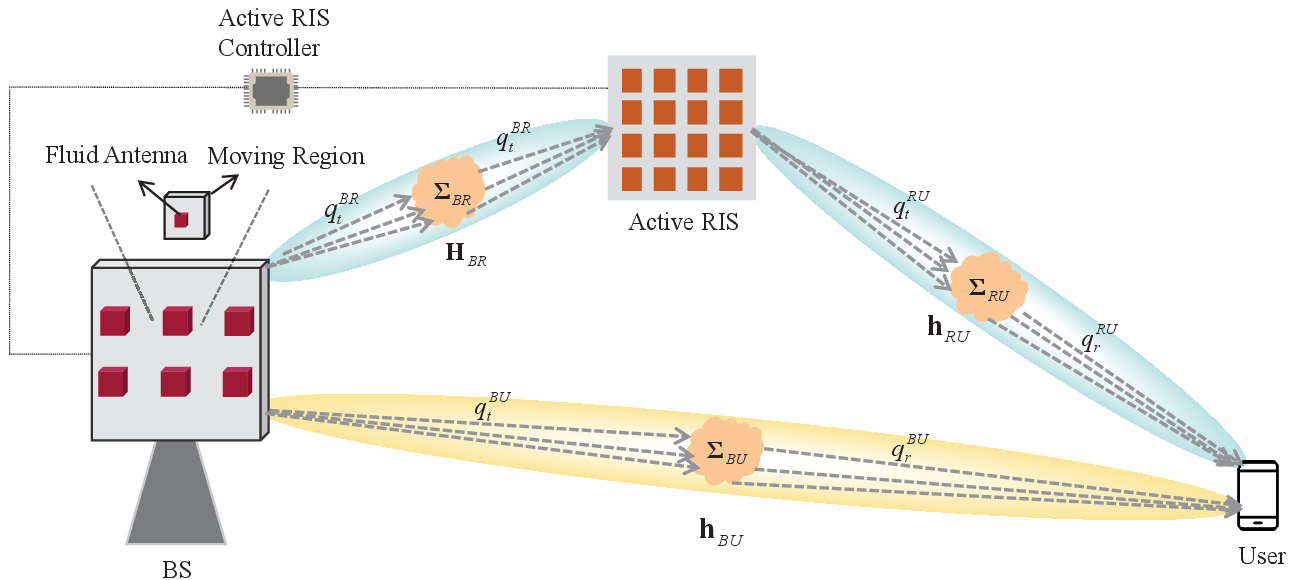}
\caption { The system model of FAS-ARIS communication systems.}
\label{model1}
\end{figure}
\subsection{Signal Model}
As shown in Fig.~\ref{model1}, we consider a FAS-ARIS downlink communication system comprising a BS, a  UE, and an ARIS. The BS is equipped $N$ FAs, and UE has an FPA. Traditional communication systems assisted by passive RIS often suffer from severe attenuation due to multiple reflections. To address this issue, we deploy an ARIS with $M$ reflecting elements, where each element provides additional signal amplification compared to traditional passive RIS. The corresponding reflection coefficient matrix is represented by  $\mathbf{E} = \mathrm{diag}\left(e_1,e_2,\cdots,e_M \right)\in\mathbb{C}^{M \times M}$, where $e_m=\beta_m\exp(j\theta_m)$ for $m\in\mathcal{M}=\{1,\cdots,M\}$, and $\beta_m>0$ and $\theta_m\in[0,2\pi]$ denote the amplification factor and the phase shift of the $m$th element in the ARIS, respectively.

Within the considered system, we assume that the transmitted downlink signal from the BS is given by \begin{align}\label{0011}
{\bf x=w}s\in\mathbb{C}^{N\times 1},
\end{align}
where $\mathbf{w}\in\mathbb{C}^{N\times 1}$ represents the transmit beamforming vector, and $s\in\mathbb{C}^{1\times 1}$ is a data stream following a complex Gaussian distribution with a mean of zero and unit variance of, i.e., $s\sim\mathcal{CN}(0,1)$.
Then, let us denote the channels between the BS and  ARIS, the BS and UE, and the ARIS and UE as $\mathbf{H}_\text{BR}\in\mathbb{C}^{M \times N}$, $\mathbf{h}_\text{BU}\in\mathbb{C}^{1 \times N}$ and $\mathbf{h}_\text{RU}\in\mathbb{C}^{1   \times M}$, respectively. The details of modeling these channels will be provided in the following subsection. Accordingly, the signal received by the UE is written as
\begin{align}\label{001}
y = \mathbf{h}_\text{BU}\mathbf{x}+\mathbf{h}_\text{RU}\mathbf{E}(\mathbf{H}_\text{BR}\mathbf{x}+\mathbf{n}_\text{R})+n_\text{u},
\end{align}
where $\mathbf{n}_\text{R} \sim\mathcal{CN}(\mathbf{0},\sigma^{2}_\text{R}\mathbf{I}_M)$ and $n_\text{u} \sim\mathcal{CN}(0,\sigma^{2}_\text{u})$ represent the Gaussian white noise generated by the infrared amplifier at the ARIS and  the UE, respectively.

\subsection{Channel Model}
Initially, we assume that the mobility region of the FAs is substantially smaller than the distance between the transmitter and the receiver. This assumption allows us to utilize the far-field communication model. Under these conditions, the amplitude differences of the signals received by each array element are negligible, and the signals exhibit a simple time-delay relationship. Additionally, the angles of departure (AoDs) and angles of arrival (AoAs) for each channel path are predominantly influenced by the scatterers and the propagation environment, remaining invariant with respect to the antenna position \cite{WMa23, LZhu23}. Specifically, we define $\mathbf{\overline{t}} = [\mathbf{t}_1, \mathbf{t}_2, \dots, \mathbf{t}_N]$ to represent the set of FAs positions at the BS, with each position $\mathbf{t}_{n}=[x_n^t, y_n^t]^T,n\in\mathcal{N}=\{1,\cdots,N\}$ denotes the position of the $n$-th FAs. Additionally, we assume that all FAs can move freely within a specified range, denoted as $S_t$.

Then, let us denote the elevation and azimuth angles at the transmit side and receive side as $\theta_{t,s_i}^i,\varphi_{t,s_i}^i ,1 \leq s_i \leq L_t^i$, and $\theta_{r,k_i}^i,\varphi_{r,k_i}^i ,1 \leq k_i \leq L_r^i, i\in\{\text{BR},\text{BU},\text{RU}\}$, respectively. Here, $L_t^i$ represents the number of transmission paths, and $L_r^i$ represents the number of reception paths. Building upon this, in $s_i$-th path, we define the signal propagation difference between the $n$-th FA and the origin $O_t$ of the transmission region as $\rho_{t,s_i}^i(\mathbf{t}_n) = x_{n}^{t}\sin\theta_{t,s_i}^i \cos\varphi_{t,s_i}^i + y_{n}^{t} \cos\theta_{t,s_i}^i, i\in\{\text{BR},\text{BU},\text{RU}\}$. Accordingly, the transmit field response vectors of a single transmit FA in the BS-ARIS links and the BS-UE links can be expressed as follow \cite{YYe24}
\begin{align}\label{002}
\bm{\zeta}_i(\mathbf{t}_n)= \left [e^{j \frac{2\pi}{\lambda}\rho_{t,1}^i(\mathbf{t}_n)}, \cdots, e^{j \frac{2\pi}{\lambda}\rho_{t,L_t^i}^i(\mathbf{t}_n)}\right]^T\in\mathbb{C}^{L_t^i},
\end{align}
for $i\in\{\text{BR},\text{BU}\}$, where $\lambda$ is the carrier wavelength. Subsequently, the transmit field response matrix of \( N \) transmitting FAs in the BS-ARIS links and the BS-UE links can be expressed as
\begin{align}\label{003}
\bm{\Upsilon}_i\mathbf{(\overline{t})} = \left [\bm{\zeta}_i(\mathbf{t}_1),\bm{\zeta}_i(\mathbf{t}_2), \cdots, \bm{\zeta}_i(\mathbf{t}_N)\right], i\in\{\text{BR},\text{BU}\}.
\end{align}

For the receive field response vector of a single reflecting element at ARIS is given by
\begin{align}\label{004}
\mathbf{f}_{\text{BR}}(\mathbf{r}_m) = \left [e^{j \frac{2\pi}{\lambda}\rho_{r,1}^{\text{BR}}(\mathbf{r}_m)},\cdots,e^{j \frac{2\pi}{\lambda}\rho_{r,L_r^{\text{BR}}}^{\text{BR}}(\mathbf{r}_m)}\right]^T \in\mathbb{C}^{L_r^{\text{BR}}},
\end{align}
where $\rho_{r,k_{\text{BR}}}^{\text{BR}}(\mathbf{r}_m) = x_m^{r}\sin\theta_{r,k_{\text{BR}}}^{\text{BR}} \cos\varphi_{r,k_{\text{BR}}}^{\text{BR}} + y_m^{r} \cos\theta_{r,k_{\text{BR}}}^{\text{BR}}$ represents the signal propagation difference of the $k_{\text{BR}}$-th path between a single reflecting element and its reception origin $O_r$. Thus, the receive field response matrix at the receiver can be expressed as
\begin{align}\label{005}
\mathbf{F}_{\text{BR}}\mathbf{(\overline{r})} = \left [\mathbf{f}_{\text{BR}}(\mathbf{r}_1),\mathbf{f}_{\text{BR}}(\mathbf{r}_2), \cdots, \mathbf{f}_{\text{BR}}(\mathbf{r}_M)\right]\in\mathbb{c}^{L_r^{\text{BR}}\times M}.
\end{align}

Because the receive positions and the transmit positions of the reflect elements at ARIS are same, we can obtain that the transmit field response matrix in the BS-ARIS links as follows

\begin{align}\label{005-1}
\bm{\Upsilon}_{\text{RU}}\mathbf{(\overline{r})} = \left [\bm{\zeta}_{\text{RU}}(\mathbf{r}_1),\bm{\zeta}_{\text{RU}}(\mathbf{r}_2), \cdots, \bm{\zeta}_{\text{RU}}(\mathbf{r}_M)\right],
\end{align}
where
\begin{align}\label{005-2}
\bm{\zeta}_{\text{RU}}(\mathbf{r}_m)= \left [e^{j \frac{2\pi}{\lambda}\rho_{t,1}^{\text{RU}}(\mathbf{r}_m)}, \cdots, e^{j \frac{2\pi}{\lambda}\rho_{t,L_t^{\text{RU}}}^{\text{RU}}(\mathbf{r}_m)}\right]^T\in\mathbb{C}^{L_t^{\text{RU}}}.
\end{align}
Furthermore, we define the path response matrix $\bm{\Sigma}_i\in\mathbb{C}^{{L_r^i}\times {L_t^i}}, i\in\{\text{BR},\text{BU},\text{RU}\}$ as the response of all transmit and receive paths between the transmitter and the receiver. Accordingly, we can represent the channels between the BS and   ARIS, the BS and UE, the ARIS and UE as
\begin{align}
\mathbf{H}_\text{BR}=&\mathbf{F}_\text{BR}^H\mathbf{(\overline{r})}\bm{\Sigma}_\text{BR}\bm{\Upsilon}_\text{BR}\mathbf{(\overline{t})}\in\mathbb{C}^{M\times N},\label{006a} \\
\mathbf{h}_\text{RU}=&\mathbf{1}^H\bm{\Sigma}_\text{RU}\bm{\Upsilon}_\text{RU}\mathbf{(\overline{r})}\in\mathbb{C}^{1\times M}, \label{006b}\\
\mathbf{h}_\text{BU}=&\mathbf{1}^H\bm{\Sigma}_\text{BU}\bm{\Upsilon}_\text{BU}\mathbf{(\overline{t})}\in\mathbb{C}^{1\times N}, \label{006c}
\end{align}
respectively.

\subsection{Problem Formulation}
Based on the channel and signal model from the above subsections,  the achievable rate from the BS to UE  can be formulated as
\begin{align}\label{008}
R=\log_2\left(1+\frac{|(\mathbf{h}_\text{RU}\mathbf{E}\mathbf{H}_\text{BR}+\mathbf{h}_\text{BU})\mathbf{w}|^2}{\sigma^{2}_\text{R}||\mathbf{h}_\text{RU}\mathbf{E}||^2+\sigma^2_\text{u}}\right).
\end{align}
Consequently,  the amplification power of the ARIS can be expressed as
\begin{align}\label{009}
\mathbb{E}\{||\mathbf{E}\left(\mathbf{H}_\text{BR}\mathbf{x}+\mathbf{n}_\text{R}    \right)||_2^2    \}=||\mathbf{E}\mathbf{H}_\text{BR}\mathbf{w}||_2^2+\sigma_\text{R}^2||\mathbf{E}||^2_F.
\end{align}
Accordingly, this paper aims at maximizing the achievable rate by jointly optimizing the transmit beamforming vector, the reflection coefficient matrix $\mathbf{E}$ at the ARIS, and the positions of the FAs at the BS $\mathbf{\overline{t}}$. Hence, we can formulate the related  optimization problem as
\begin{subequations}\label{010}
\begin{align}
\max\limits_{\mathbf{\overline{t}},\mathbf{w},\mathbf{E}} \quad &R \label{010a}\\
\mathrm{s.t.} \quad \ &\mathbf{\overline{t}} \in S_t, \label{010b}\\
&||\mathbf{t}_n-\mathbf{t}_v||_2\geq D,~n,v\in\mathcal{N},~n\neq v, \label{010c}\\
&||\mathbf{w}||_2^2\leq P_0, \label{010d}\\
&||\mathbf{E}\mathbf{H}_\text{BR}\mathbf{w}||_2^2+\sigma_\text{R}^2||\mathbf{E}||^2_F\leq P_1, \label{010f}
\end{align}
\end{subequations}
where $D$ is the minimum distance between adjacent antennas to prevent coupling; $P_0$ is the maximum transmit power at the BS; $P_1$ is the maximum transmit power at the ARIS. However,
due to the coupling of optimization variables within both the objective function and the constraints, along with the highly non-convex nature of Problem \eqref{010}, addressing this issue poses significant challenges.
\section{Proposed Algorithm}
In this section, to address the significant challenges of Problem \eqref{010}, we decompose it into three sub-problems concerning  the BS beamforming vector $\mathbf{w}$, the positions of the FAs $\mathbf{\overline{t}}$, and the reflection coefficient matrix $\mathbf{E}$. We then apply the AO algorithm to iteratively solve these sub-problems, ultimately achieving a locally optimal solution.

\subsection{BS's Beamforming Optimization}
As the first step of the employed AO algorithm, our objective is to optimize $\mathbf{w}$ with given $\mathbf{\overline{t}}$ and $\mathbf{E}$. Accordingly, we can reformulate Problem \eqref{010} as
\begin{subequations}\label{0011}
\begin{align}
\max\limits_{\mathbf{w}} \quad &R \label{0011a}\\
\mathrm{s.t.} \quad \ &||\mathbf{w}||_2^2\leq P_0, \label{0011b}\\
&||\mathbf{E}\mathbf{H}_\text{BR}\mathbf{w}||_2^2+\sigma_\text{R}^2||\mathbf{E}||^2_F\leq P_1.\label{0011d}
\end{align}
\end{subequations}
 Then, due to the fact  that $\log_2(1 + x)$ is an increasing function with respect to $x$, maximizing $R$ can be regarded as maximizing $|(\mathbf{h}_\text{RU}\mathbf{E}\mathbf{H}_\text{BR}+\mathbf{h}_\text{BU})\mathbf{w}|^2$. Then, by letting $\bm{\varpi}=\mathbf{h}_\text{RU}\mathbf{E}\mathbf{H}_\text{BR}+\mathbf{h}_\text{BU}$ and $\mathbf{W}=\mathbf{w}\mathbf{w}^H$, Problem \eqref{0011} can be further reformulated as
\begin{subequations}\label{011}
\begin{align}
\max\limits_{\mathbf{W}\succeq\mathbf{0}} \quad &\bm{\varpi}\mathbf{W}\bm{\varpi}^H \label{011a}\\
\mathrm{s.t.} \quad \ &\mathrm{Tr}(\mathbf{W})\leq P_0, \label{011b}\\
&\mathrm{Tr}(\mathbf{E}\mathbf{H}_\text{BR}\mathbf{W}\mathbf{H}_\text{BR}^H\mathbf{E}^H)+\sigma_\text{R}^2\mathrm{Tr}(\mathbf{E}\mathbf{E}^H)\leq P_1, \label{011c}\\
&\mathrm{rank}(\mathbf{W})=1. \label{011d}
\end{align}
\end{subequations}
It can be observed that by removing the rank-one constraint, Problem \eqref{011} is transformed into a standard semi-definite programming (SDP) problem, which can be solved using CVX to obtain $\mathbf{\widehat{W}}$ \cite{Boyd}. If $\mathrm{rank}(\mathbf{\widehat{W}})=1$, we can get the optimal $\mathbf{w}$, i.e., $\mathbf{w}^o$ by the eigenvalue decomposition. If $\mathrm{rank}(\mathbf{\widehat{W}})>1$, the rank-one solution can be derived as follows:
\begin{align}\label{012-1}
\mathbf{W}=\mathbf{w}^o{\mathbf{w}^o}^H,
\end{align}
where
\begin{align}\label{012-2}
\mathbf{w}^o=(\bm{\varpi}\mathbf{W}\bm{\varpi}^H)^{-\frac{1}{2}}\mathbf{W}\bm{\varpi}^H,
\end{align}
whose the proof is shown in Appendix \ref{A1}.

\subsection{FAs' Position Optimization}
As the second sub-problem of Problem \eqref{010} when employing AO algorithm, our objective is to optimize the positions $\mathbf{\overline{t}}$ of the FAs with given $\mathbf{W}$ and $\mathbf{E}$. Accordingly, Problem \eqref{010} can be rewritten as
\begin{subequations}\label{0013}
\begin{align}
\max\limits_{\mathbf{\overline{t}}} \quad &R\label{0013a}\\
\mathrm{s.t.} \quad \ &\mathbf{\overline{t}} \in S_t, \label{0013b}\\
&||\mathbf{t}_n-\mathbf{t}_v||_2\geq D,~n,v\in\mathcal{N},~n\neq v, \label{0013c}\\
&\mathrm{Tr}\left(\mathbf{E}\mathbf{H}_\text{BR}\mathbf{W}\mathbf{H}_\text{BR}^H\mathbf{E}^H  \right)+\sigma_\text{R}^2\mathrm{Tr}\left(\mathbf{E}\mathbf{E}^H  \right)\leq P_1. \label{0013d}
\end{align}
\end{subequations}
Then, by applying the monotonicity of $\log_2(1 + x)$,  we can rewrite the optimization problem as
\begin{subequations}\label{013}
\begin{align}
\max\limits_{\mathbf{\overline{t}}} \quad &(\mathbf{h}_\text{RU}\mathbf{E}\mathbf{H}_\text{BR}+\mathbf{h}_\text{BU})\mathbf{W}(\mathbf{h}_\text{RU}\mathbf{E}\mathbf{H}_\text{BR}+\mathbf{h}_\text{BU})^H \label{013a}\\
\mathrm{s.t.} \quad \ &\mathbf{\overline{t}} \in S_t, \label{013b}\\
&||\mathbf{t}_n-\mathbf{t}_v||_2\geq D,~n,v\in\mathcal{N},~n\neq v, \label{013c}\\
&\mathrm{Tr}\left(\mathbf{E}\mathbf{H}_\text{BR}\mathbf{W}\mathbf{H}_\text{BR}^H\mathbf{E}^H  \right)+\sigma_\text{R}^2\mathrm{Tr}\left(\mathbf{E}\mathbf{E}^H  \right)\leq P_1. \label{013d}
\end{align}
\end{subequations}
Building upon this,  the objective function \eqref{013a} is then transformed into a more tractable form as
\begin{align}\label{014}
&(\mathbf{h}_\text{RU}\mathbf{E}\mathbf{H}_\text{BR}+\mathbf{h}_\text{BU})\mathbf{W}(\mathbf{h}_\text{RU}\mathbf{E}\mathbf{H}_\text{BR}+\mathbf{h}_\text{BU})^H \nonumber \\
&=\left(\bm{\xi}^H\bm{\Upsilon}_\text{BR}\mathbf{(\overline{t})}+\bm{\omega}^H\bm{\Upsilon}_\text{BU}\mathbf{(\overline{t})}  \right)\mathbf{W}\left(\bm{\Upsilon}_\text{BR}^H\mathbf{(\overline{t})}\bm{\xi}+\bm{\Upsilon}_\text{BU}^H\mathbf{(\overline{t})}\bm{\omega}  \right),
\end{align}
where $\bm{\xi}^H=\mathbf{h}_\text{RU}\mathbf{E}\mathbf{F}_\text{BR}^H\mathbf{(\overline{r})}\bm{\Sigma}_\text{BR}$ and $\bm{\omega}^H=\mathbf{1}^H\bm{\Sigma}_\text{BU}$ are invariant
to $\mathbf{\overline{t}}$. By define ${\mu}_n=\bm{\xi}^H\bm{\zeta}_\text{BR}(\mathbf{t}_n)+\bm{\omega}^H\bm{\zeta}_\text{BR}(\mathbf{t}_n), n\in\mathcal{N}$, Eq. \eqref{014} can be further written as
\begin{align}\label{015}
\underbrace{\left[\mathbf{W} \right]_{n,n}|{\mu}_n|^2+2\Re\{\widetilde{\alpha}{\mu}_n\}}_{{{g}}(\mathbf{t}_n)}+\widetilde{\beta},
\end{align}
where
\begin{align}\label{016}
&~~~~~~~~~~~~~~~~~~~\widetilde{\alpha}=\sum_{l\neq n}^{N}\left[\mathbf{W} \right]_{n,l}{\mu}_l^{*},\\
&\widetilde{\beta}=\sum_{l\neq n}^{N}\left(\sum_{j\neq n}^{l-1} 2\Re\{\left[\mathbf{W}\right]_{l,j} \mu_l \mu_j^{*} + \left[\mathbf{W}\right]_{l,l}|\mu_l|^2    \}    \right).
\end{align}
Given that $\{\mathbf{t}_l,l\neq n\}_{l=1}^{N}$ is known, $\widetilde{\alpha},\widetilde{\beta}$ and $\mu_l$ remain invariant to $\mathbf{t}_n$. Therefore, maximizing \eqref{013a} is equivalent to maximizing $g(\mathbf{t}_n)$. However, although $g(\mathbf{t}_n)$ is a linear function of $\bm{\zeta}_\text{BR}(\mathbf{t}_n)$, it is non-convex and non-concave with respect to $\mathbf{t}_m$. To address this issue, we employ the second-order Taylor expansion to construct a surrogate function that locally approximates the objective function. We denote the gradient vector and the Hessian matrix of $g(\mathbf{t}_n)$ as $\nabla g(\mathbf{t}_n)$ and $\nabla^2 g(\mathbf{t}_n)$, respectively, with detailed derivations provided in Appendix \ref{A2}. Then, we introduce a scalar $\kappa_n$ such that $\kappa_n \mathbf{I}_2\succeq \nabla^2 g(\mathbf{t}_n)$, with the specific expression detailed in Appendix  \ref{A3}.  At this juncture, we can determine the global lower bound of $ g(\mathbf{t}_n)$ as follow
\begin{align}\label{018}
g(\mathbf{t}_n)\geq &g(\mathbf{t}_n^{(q)})+\nabla g(\mathbf{t}_n^{(q)})^T\left(\mathbf{t}_n-\mathbf{t}_n^{(q)}   \right) \nonumber\\
&-\frac{\kappa_n}{2}\left(\mathbf{t}_n-\mathbf{t}_n^{(q)}   \right)^T\left(\mathbf{t}_n-\mathbf{t}_n^{(q)}   \right) \nonumber\\
=&{{-\frac{\kappa_n}{2}\mathbf{t}_n^T\mathbf{t}_n+\left(\nabla g(\mathbf{t}_n^{(q)})+\kappa_n\mathbf{t}_n \right)^T\mathbf{t}_n}} \nonumber\\
&+\underbrace{g(\mathbf{t}_n^{(q)})-\frac{\kappa_n}{2}(\mathbf{t}_n^{(q)})^T(\mathbf{t}_n^{(q)})-\nabla g(\mathbf{t}_n^{(q)})^T\mathbf{t}_n^{(q)}}_\text{constant},
\end{align}
where $\mathbf{t}_n^{(q)}$ is the value of $\mathbf{t}_n$ at the $q$-th internal iteration.

Next, we address the non-convex constraint  \eqref{013d}. For the first term on the left side of \eqref{013d}, we reformulate it as follows:
\begin{align}\label{019}
\mathrm{Tr}\left(\mathbf{E}\mathbf{H}_\text{BR}\mathbf{W}\mathbf{H}_\text{BR}^H\mathbf{E}^H  \right)&=\mathrm{Tr}\left(\mathbf{H}_\text{BR}^H\mathbf{E}^H\mathbf{E}\mathbf{H}_\text{BR}\mathbf{W}  \right) \nonumber \\
&=\mathrm{Tr}\left(\bm{\Upsilon}_\text{BR}\mathbf{(\overline{t})}^H \bm{\Psi} \bm{\Upsilon}_\text{BR}\mathbf{(\overline{t})} \mathbf{W}    \right)  \nonumber \\
&=\underbrace{\mathrm{Tr}\left(\bm{\Upsilon}_\text{BR}\mathbf{(\overline{t})} \mathbf{W}\bm{\Upsilon}_\text{BR}\mathbf{(\overline{t})}^H \bm{\Psi} \right)}_{\overline{g}(\mathbf{t}_n)}
\end{align}
where $\bm{\Psi}=\bm{\Sigma}_{\text{BR}}^H \mathbf{F}_{\text{BR}}\mathbf{(\overline{r})}\mathbf{E}^H \mathbf{E} \mathbf{F}_{\text{BR}}^H \mathbf{(\overline{r})} \bm{\Sigma}_{\text{BR}}$. Similarly, since our optimization goal is the position $\mathbf{t}_n$ of the $n$-th antenna, we can expand \eqref{019} into the following form
\begin{align}\label{020}
{\overline{g}(\mathbf{t}_n)}=\widetilde{g}(\mathbf{t}_n)+2\Re\{\bm{\zeta}_{\text{BR}}^H(\mathbf{t}_n) \bm{\Psi}\bm{\tau}  \}+c_1,
\end{align}
where $\mathbf{w}(n)$ is the $n$-th element of $\mathbf{w}$ obtained from Problem \eqref{011}. The detailed derivation process is provided in Eq. \eqref{021} at the top of the next page.

\begin{figure*}[tb]
\centering
\begin{align}
\label{021}
{\overline{g}(\mathbf{t}_n)}=&\mathrm{Tr}\left(\left[\sum_{n=1}^{N}\bm{\zeta}_\text{BR}(\mathbf{t}_n)\mathbf{w}(n) \right]\left[\sum_{n=1}^{N}\mathbf{w}^H(n)\bm{\zeta}_\text{BR}^H(\mathbf{t}_n)  \right]\bm{\Psi} \right)\nonumber\\
=&\mathrm{Tr}\left(\underbrace{\mathbf{w}(n)\mathbf{w}^H(n)\bm{\zeta}_\text{BR}(\mathbf{t}_n)
\bm{\zeta}_\text{BR}^H(\mathbf{t}_n)\bm{\Psi}}_{\widetilde{g}(\mathbf{t}_n)}  \right)+\mathrm{Tr}\left(\bm{\zeta}_\text{BR}(\mathbf{t}_n)\mathbf{w}(n)\sum_{l\neq n}^{N} \mathbf{w}^H(l)\bm{\zeta}_\text{BR}^H(\mathbf{t}_l)\bm{\Psi}   \right)\nonumber\\
&+\mathrm{Tr}\left(\underbrace{\sum_{j\neq n}^{N} \bm{\zeta}_\text{BR}(\mathbf{t}_j)\mathbf{w}(j)\mathbf{w}^H(n)}_{\bm{\tau}}\bm{\zeta}_\text{BR}^H(\mathbf{t}_n)\bm{\Psi} \right) + \underbrace{\mathrm{Tr}\left(\sum_{j\neq n}^{N} \bm{\zeta}_\text{BR}(\mathbf{t}_j)\mathbf{w}(j) \sum_{l\neq n}^{N} \mathbf{w}^H(l)\bm{\zeta}_\text{BR}^H(\mathbf{t}_l)\bm{\Psi} \right),}_{c_1}
\end{align}
\centering
\hrulefill
\end{figure*}

Then, we employ the MM iterative algorithm to address this non-convex constraint. Specifically, we denote the value of ${\overline{g}(\mathbf{t}_n)}$ at the $q$-th iteration as ${\overline{g}(\mathbf{t}_n^{(q)})}$. Then, at the $(q+1)$-th iteration, we introduce  $\varepsilon(\mathbf{t}_n)$ as a surrogate function for ${\overline{g}(\mathbf{t}_n^{(q)})}$ that satisfies the following three conditions \cite{JTang2024}:
\begin{align}
\varepsilon(\mathbf{t}_n^{(q)})&={\overline{g}(\mathbf{t}_n^{(q)})}, \label{022a} \\
\nabla\varepsilon(\mathbf{t}_n^{(q)})|&=\nabla{\overline{g}(\mathbf{t}_n^{(q)})},\label{022b} \\
\varepsilon(\mathbf{t}_n) &\geq {\overline{g}(\mathbf{t}_n)}. \label{022c}
\end{align}
Conditions \eqref{022a} and \eqref{022b} ensure that the surrogate function's value and gradient at $\mathbf{t}_n^{(q)}$ match those of the original function. Meanwhile, Condition \eqref{022c} guarantees that the surrogate function serves as an upper bound to the original function. This alignment allows us to represent the original function with $\varepsilon(\mathbf{t}_n)$ at the $(q+1)$-th iteration.

According to \eqref{022a}, \eqref{022a}, and \eqref{022a}, for any value of $\mathbf{t}_n$ at the $q$-th iteration, we can construct the surrogate function as the upper bound of $\widetilde{g}(\mathbf{t}_n)$, which can be expressed as
\begin{align}\label{023}
\widetilde{g}(\mathbf{t}_n)\leq & \bm{\zeta}^H_{\text{BR}}(\mathbf{t}_n)\bm{\Phi} \bm{\zeta}_{\text{BR}}(\mathbf{t}_n) + \bm{\zeta}^H_{\text{BR}}(\mathbf{t}_n^{(q)})\left(  \bm{\Phi}-\widetilde{\bm{\Psi}} \right) \bm{\zeta}_{\text{BR}}(\mathbf{t}_n^q) \nonumber \\
&-2\Re\{\bm{\zeta}^H_{\text{BR}}(\mathbf{t}_n)\left( \bm{\Phi}-\widetilde{\bm{\Psi}} \right) \bm{\zeta}_{\text{BR}}(\mathbf{t}_n^{(q)})           \},
\end{align}
where $\widetilde{\bm{\Psi}}=\bm{\Psi}\mathbf{w}(n)\mathbf{w}^H(n)$ and $\bm{\Phi}=\lambda_\text{max}\mathbf{I}_{L_t^{\text{BR}}}$, $\lambda_\text{max}$ is the maximum eigenvalue of $\widetilde{\bm{\Psi}}$. Furthermore, we can observe that $\bm{\zeta}^H_{\text{BR}}(\mathbf{t}_n)\bm{\Phi} \bm{\zeta}_{\text{BR}}(\mathbf{t}_n)=\lambda_\text{max} L_t^{\text{BR}}$, and the second term on the right side of \eqref{023} is known. We define
\begin{align}\label{0023}
c_2=\lambda_\text{max} L_t^{\text{BR}}+\bm{\zeta}^H_{\text{BR}}(\mathbf{t}_n^{(q)})\left(  \bm{\Phi}-\widetilde{\bm{\Psi}} \right) \bm{\zeta}_{\text{BR}}(\mathbf{t}_n^{(q)}).
\end{align}
Thus, we can rewrite constraint \eqref{013d} in the following form
\begin{align}\label{024}
\underbrace{2\Re\{\bm{\zeta}^H_{\text{BR}}(\mathbf{t}_n)\bm{\eta}         \}}_{\widehat{g}(\mathbf{t}_n)}\leq \hat{P_1}
\end{align}
where
\begin{align}\label{025}
\bm{\eta}&=\bm{\Psi}\bm{\tau}-\left(  \bm{\Phi}-\widetilde{\bm{\Psi}} \right) \bm{\zeta}_{\text{BR}}(\mathbf{t}_n^{(q)}), \\
\hat{P_1}&=P_1-c_1-c_2-\sigma_\text{R}^2\mathrm{Tr}\left(\mathbf{E}\mathbf{E}^H  \right).
\end{align}
However, $\widehat{g}(\mathbf{t}_n)$ remains a non-concave function of $\mathbf{t}_n$,  precluding the use of the first-order Taylor expansion to obtain its upper bound. Therefore, as previously mentioned, we introduce a scalar $\widehat{\kappa}_n$ and employ the second-order Taylor expansion of $\widehat{g}(\mathbf{t}_n)$ to construct a surrogate function that provides the upper bound of $\widehat{g}(\mathbf{t}_n)$. Specifically, the formulation is as follows:
\begin{align}\label{026}
\widehat{g}(\mathbf{t}_n)\leq ~&\widehat{g}(\mathbf{t}_n^{(q)})+\nabla \widehat{g}(\mathbf{t}_n^{(q)})^T\left(\mathbf{t}_n-\mathbf{t}_n^{(q)}   \right) \nonumber\\
&+\frac{\widehat{\kappa}_n}{2}\left(\mathbf{t}_n-\mathbf{t}_n^{(q)}   \right)^T\left(\mathbf{t}_n-\mathbf{t}_n^{(q)}   \right) \nonumber\\
\triangleq~& \widehat{\delta}(\mathbf{t}_n),
\end{align}
where
\begin{align}
\nabla \widehat{g}(\mathbf{t}_n)&=\left[\frac{\partial \widehat{g}(\mathbf{t}_n)}{\partial x^{t}_n},\frac{\partial \widehat{g}(\mathbf{t}_n)}{\partial y^{t}_n}     \right], \label{0026a} \\
\frac{\partial \widehat{g}(\mathbf{t}_n)}{\partial x^{t}_n}&=-\frac{4\pi}{\lambda}\left( \sum\limits_{s=1}^{L_t^\text{BR}}|\bm{\eta}_s|\sin\theta_{t,s}^{\text{BR}}\cos\varphi_{t,s}^{\text{BR}}\sin\left(\nu^5_s(\mathbf{t}_n)\right)       \right), \label{0026b} \\
\frac{\partial \widehat{g}(\mathbf{t}_n)}{\partial y^{t}_n}&=-\frac{4\pi}{\lambda}\left( \sum\limits_{s=1}^{L_t^\text{BR}}|\bm{\eta}_s|\cos\theta_{t,s}^{\text{BR}}\sin\left(\nu^5_s(\mathbf{t}_n)\right)       \right), \label{0026b} \\
\nu^5_s(\mathbf{t}_n)&=\frac{2\pi}{\lambda}\rho_{t,s}^{\text{BR}}(\mathbf{t}_n)-\angle\bm{\eta}_s,  \label{0026c} \\
\widehat{\kappa}_n&=\frac{16\pi^2}{\lambda^2}\sum\limits_{s=1}^{L_t^\text{BR}}|\bm{\eta}_s|,\label{0026c}
\end{align}
whose proofs are similar to those in the proof provided in the Appendix \ref{A2} and Appendix \ref{A3}.

Furthermore, we address the non-convex constraint \eqref{013c}. It can be observed that $||\mathbf{t}_n-\mathbf{t}_v||_2$ is a concave function of $\mathbf{t}_n$, enabling us to directly obtain its lower bound through a first-order Taylor expansion at $\mathbf{t}_n^{(q)}$. Specifically, the expansion is given as
\begin{align}\label{027}
||\mathbf{t}_n-\mathbf{t}_v||_2 &\geq ||\mathbf{t}_n^{(q)}-\mathbf{t}_v||_2+\left(\nabla ||\mathbf{t}_n^{(q)}-\mathbf{t}_v||_2\right)^T \left(\mathbf{t}_n-\mathbf{t}_n^{(q)} \right) \nonumber \\
&=\frac{1}{||\mathbf{t}_n^{(q)}-\mathbf{t}_v||_2}(\mathbf{t}_n^{(q)}-\mathbf{t}_v)^T(\mathbf{t}_n-\mathbf{t}_v) \nonumber \\
&\triangleq \gamma(\mathbf{t}_n).
\end{align}

Finally, based on the above derivations, we rewrite the final optimization problem in the following form
\begin{subequations}\label{028}
\begin{align}
\max\limits_{\mathbf{t}_n} \quad &{{-\frac{\kappa_n}{2}\mathbf{t}_n^T\mathbf{t}_n+\left(\nabla g(\mathbf{t}_n^{(q)})+\kappa_n^q\mathbf{t}_n^{(q)} \right)^T\mathbf{t}_n}} \label{028a} \\
\mathrm{s.t.} \quad   &\gamma(\mathbf{t}_n)\geq D, n,v\in\mathcal{N},~n\neq v, \label{028b}\\
&\widehat{\delta}(\mathbf{t}_n) \leq \hat{P_1}, \label{028c} \\
&\eqref{013b}.
\end{align}
\end{subequations}
Here, the objective function \eqref{028a} is a concave quadratic function of $\mathbf{t}_n$, constraints \eqref{028b} and \eqref{028c} are both linear, and constraint \eqref{013b} is a concave function. This aligns with the principles of CVX, allowing us to directly use the CVX toolbox for solving the problem \cite{MGrant}.

\subsection{Reflection Coefficient Matrix  Optimization}
As the third sub-problem of Problem \eqref{010} when employing AO algorithm, we fix $\overline{\mathbf{t}}$ and $\mathbf{W}$ and optimize $\mathbf{E}$, the sub-problem can be rewritten as follow
\begin{subequations}\label{029}
\begin{align}
\max\limits_{\mathbf{E}} \quad &\frac{|(\mathbf{h}_\text{RU}\mathbf{E}\mathbf{H}_\text{BR}+\mathbf{h}_\text{BU})\mathbf{w}|^2}{\sigma^{2}_\text{R}||\mathbf{h}_\text{RU}\mathbf{E}||^2+\sigma^2_\text{u}} \label{029a}\\
\mathrm{s.t.} \quad
&||\mathbf{E}\mathbf{H}_\text{BR}\mathbf{w}||_2^2+\sigma_\text{R}^2||\mathbf{E}||^2_F\leq P_1. \label{029b}
\end{align}
\end{subequations}
Let us define $\mathbf{e}=[e_1,e_2, \cdots e_M]^H$ as the vector consisting of the diagonal elements of $\mathbf{E}$ and $\widetilde{\mathbf{e}}=[\mathbf{e}^H,1]^H$ as the vector containing $\mathbf{e}$ and $1$. Furthermore, since
\begin{align}
\mathbf{h}_\text{RU}\mathbf{E}\mathbf{H}_\text{BR}\mathbf{w}&=\mathbf{e}^H \mathrm{diag}\left( \mathbf{h}_{\text{RU}}\right)\mathbf{H}_{\text{BR}}\mathbf{w}, \label{030a} \\
\mathbf{h}_\text{RU}\mathbf{E}&=\mathbf{e}^H \mathrm{diag}\left( \mathbf{h}_{\text{RU}}\right), \label{030b} \\
\mathbf{E}\mathbf{H}_\text{BR}\mathbf{w}&=\mathbf{e}^H \mathrm{diag}\left( \mathbf{H}_{\text{BR}}\mathbf{w}\right). \label{030c}
\end{align}
We can transform Problem \eqref{029} into the following form
\begin{subequations}\label{031}
\begin{align}
\max\limits_{\widetilde{\mathbf{E}}\succeq 0} \quad &\frac{\mathrm{Tr}(\mathbf{V}\widetilde{\mathbf{E}})+\mathbf{h}_\text{BU}\mathbf{w}\mathbf{w}^H\mathbf{h}^H_\text{BU}}{\mathrm{Tr}(\overline{\mathbf{V}}\widetilde{\mathbf{E}})} \label{031a}\\
\mathrm{s.t.} \quad &\mathrm{Tr}(\widehat{\mathbf{V}}\widetilde{\mathbf{E}})\leq P_1, \label{031b}\\
&|\mathrm{diag}(\widetilde{\mathbf{E}})|_{M+1}=1, \label{031c} \\
&\mathrm{rank}(\widetilde{\mathbf{E}})=1, \label{031d}
\end{align}
\end{subequations}
where $\widetilde{\mathbf{E}}=\widetilde{\mathbf{e}}\widetilde{\mathbf{e}}^H$ and the expressions for $\mathbf{V}$, $\overline{\mathbf{V}}$ and $\widehat{\mathbf{V}}$ can be found in \eqref{0031a}, \eqref{0031b} and \eqref{0031c}.

\begin{figure*}[t]
\centering
\begin{align}
{\mathbf{V}}&=
\begin{bmatrix}
\mathrm{diag}\left( \mathbf{h}_{\text{RU}}\right)\mathbf{H}_{\text{BR}}\mathbf{w}\mathbf{w}^H\mathbf{H}_{\text{BR}}^H\mathrm{diag}\left( \mathbf{h}^H_{\text{RU}}\right)& \mathrm{diag}\left( \mathbf{h}_{\text{RU}}\right)\mathbf{H}_{\text{BR}}\mathbf{w}\mathbf{w}^H\mathbf{h}_{\text{BU}} \\
\mathbf{h}_{\text{BU}}\mathbf{w}\mathbf{w}^H\mathbf{H}^H_{\text{BR}}\mathrm{diag}\left( \mathbf{h}^H_{\text{RU}}\right) & 0
\end{bmatrix}, \label{0031a}  \\
\overline{\mathbf{V}}&=
\begin{bmatrix}
\sigma^{2}_\text{R}\mathrm{diag}\left( \mathbf{h}_{\text{RU}}\right)\mathrm{diag}\left( \mathbf{h}^H_{\text{RU}}\right) & 0 \\
0 & \sigma^2_\text{u}
\end{bmatrix}, \label{0031b}  \\
\widehat{\mathbf{V}}&=
\begin{bmatrix}
\mathrm{diag}\left( \mathbf{H}_{\text{BR}}\mathbf{w}\right)\mathrm{diag}\left( \mathbf{w}^H\mathbf{H}^H_{\text{BR}}\right)+ \sigma_\text{R}^2\mathbf{I} & 0 \\
0 & 0
\end{bmatrix}. \label{0031c}
\end{align}
\centering
\hrulefill
\end{figure*}

Next, we address Problem \eqref{029}. First,  for the non-convex constraint \eqref{031d}, we adopt the sequential rank-one constraint relaxation (SRCR) method, equivalently transforming the rank-one constraint into \cite{JZou23}
\begin{align}\label{032}
\mathbf{u}_\text{max}^H\left( \widetilde{\mathbf{E}}^{(p)}\right)\widetilde{\mathbf{E}}\mathbf{u}_\text{max}\left( \widetilde{\mathbf{E}}^{(p)}\right)\geq \vartheta^{(p)}\mathrm{Tr}(\widetilde{\mathbf{E}}),
\end{align}
where $\widetilde{\mathbf{E}}^{(p)}$ is the value at the $p$-th internal iteration, $\mathbf{u}_\text{max}\left( \widetilde{\mathbf{E}}^{(p)}\right)$ is the eigenvector corresponding to the largest eigenvalue $\lambda_\text{max}\left( \widetilde{\mathbf{E}}^{(p)} \right)$ of $\widetilde{\mathbf{E}}^{(p)}$ and $\vartheta^{(p)}$ represents the relaxation parameter and updated by following
\begin{align}\label{0032}
\vartheta^{(p+1)} = \min\left(1,\frac{\lambda_\text{max}\left( \widetilde{\mathbf{E}}^{(p+1)}\right)}{\mathrm{Tr}(\widetilde{\mathbf{E}}^{(p+1)})} + \varepsilon^{(p+1)} \right),
\end{align}
where $o^{(p+1)}$ is a given scalar.

However, Problem \eqref{031} remains a complex fractional programming problem, which is challenging to solve. To reduce the complexity, we first introduce auxiliary variables to equivalently transform the problem into a more tractable form. Then, we use the successive convex approximation (SCA) algorithm to iteratively approximate these non-convex functions with linear functions, thereby solving the problem. Specifically, by introducing auxiliary variables $\chi$ and $\varrho$, we transform Problem \eqref{031} into the following form
\begin{subequations}\label{033}
\begin{align}
\max\limits_{\chi,\varrho,\widetilde{\mathbf{E}}} \quad &\chi \label{033a}\\
\mathrm{s.t.} \quad &\mathrm{Tr}(\mathbf{V}\widetilde{\mathbf{E}})+\mathbf{h}_\text{BU}\mathbf{w}\mathbf{w}^H\mathbf{h}^H_\text{BU}\geq \chi\varrho, \label{033b} \\
&\mathrm{Tr}(\overline{\mathbf{V}}\widetilde{\mathbf{E}})\leq \varrho, \label{033c} \\
&\mathrm{Tr}(\widehat{\mathbf{V}}\widetilde{\mathbf{E}})\leq P_1, \label{033d}\\
&|\mathrm{diag}(\widetilde{\mathbf{E}})|_{M+1}=1, \label{033e} \\
&\eqref{032}.
\end{align}
\end{subequations}

It is worth noting that constraint \eqref{033b} is still non-convex. For this reason, we propose an SCA algorithm to find the a upper bound of $\chi\varrho$. Specifically , let
\begin{align}\label{034}
\chi\varrho=\frac{1}{4}\left(\chi+\varrho\right)^2-\frac{1}{4}\left(\chi-\varrho\right)^2.
\end{align}
It can be seen that $\frac{1}{4}\left(\chi+\varrho\right)^2$ and $\frac{1}{4}\left(\chi-\varrho\right)^2$ are convex on both $\chi$ and $\varrho$. Thus, we have
\begin{align}\label{035}
\chi\varrho\leq & \frac{1}{4}\left(\chi+\varrho\right)^2-\frac{1}{4}\left(\chi^{(p)}+\varrho^{(p)}\right)^2 \nonumber \\
&-\frac{1}{2}\left(\chi^{(p)}+\varrho^{(p)}\right)\left(\chi-\chi^{(p)}-\varrho+\varrho^{(p)}     \right) \nonumber \\
\triangleq & f(\chi, \varrho;\chi^{(p)}, \varrho^{(p)}).
\end{align}
Based on the above analysis, we can build the final optimization problem
\begin{subequations}\label{036}
\begin{align}
\max\limits_{\chi, \varrho, \widetilde{\mathbf{E}}} \quad &\chi \label{036a}\\
\mathrm{s.t.} \quad &\mathrm{Tr}(\mathbf{V}\widetilde{\mathbf{E}})+\mathbf{h}_\text{BU}\mathbf{w}\mathbf{w}^H\mathbf{h}^H_\text{BU}\geq f(\chi, \varrho;\chi^{(p)}, \varrho^{(p)}), \label{036b} \\
&\eqref{033c},\eqref{033d},\eqref{033e},\eqref{032}.
\end{align}
\end{subequations}
Therefore, Problem \eqref{036} is a standard SDP problem that can be solved directly using the CVX toolbox and summarize the process in Algorithm 1.

Finally, we summarize the whole process of the AO algorithm in Algorithm 2.

\begin{algorithm}[h]
\caption{The Proposed SCA-Based SRCR Algorithm for Solving Problem \eqref{029}}
\begin{algorithmic}[1]
\STATE \textbf{Initialize:} $\chi^{(0)}$, $\varrho^{(0)}$, $\vartheta^{(0)}$, $\varepsilon^{(0)}$.
\STATE \textbf{while} Increase of the  transmission rate R is above $\epsilon_2$\\
\STATE \textbf{do} \\
\STATE \textbf{if}  If Problem \eqref{036} has a solution, then by solving Problem \eqref{036}, we update $\widetilde{\mathbf{E}}^{(p+1)}$ and letting $\varepsilon^{(p+1)}=\varepsilon^0$ \\
\STATE \textbf{else} \\
\STATE \quad Update $\widetilde{\mathbf{E}}^{(p+1)}=\widetilde{\mathbf{E}}^{(p)}$;  \\
\STATE \quad Update $\varepsilon^{(p+1)}=\varepsilon^{(p)}/2$;  \\
\STATE \textbf{end if} \\
\STATE \quad Update $\vartheta^{(p+1)}$ by \eqref{0032};
\STATE \quad $p:=p+1$;         \\
\STATE \textbf{end while}
\STATE Obtain $\mathbf{E}$ by decompose $\widetilde{\mathbf{E}}$.
\end{algorithmic}
\end{algorithm}

\begin{algorithm}[t]
\caption{The Proposed Alternating Optimization for Solving Problem \eqref{010}}
\begin{algorithmic}[1]
\STATE \textbf{Initialize:} $\mathbf{\overline{t}}^{(0)}$ and $\mathbf{E}^{(0)}$.
\STATE \textbf{while} Increase of the  transmission rate R is above $\epsilon_3$\\
\STATE \textbf{do} \\
\STATE \quad Obtain $\mathbf{W}$ by solving Problem \eqref{011} and update $\mathbf{w}$; \\
\STATE \quad \textbf{while} Increase of the  transmission rate R is above $\epsilon_1$\\
\STATE \quad \textbf{do} \\
\STATE \quad \quad \textbf{for} $n=1\rightarrow N$ \textbf{do} \\
\STATE \quad \quad \quad Update $\mathbf{\overline{t}}$ by solving Problem \eqref{028}; \\
\STATE \quad \quad \textbf{end}
\STATE \quad \textbf{end while}
\STATE \quad Update $\mathbf{E}$ by solving Problem \eqref{036}; \\
\STATE \textbf{end while}
\STATE Obtain the optimal solutions for $\mathbf{W}$, $\mathbf{\overline{t}}$ and $\mathbf{E}$.
\end{algorithmic}
\end{algorithm}

\section{Numerical Result}
In this section, we provide numerical results to evaluate the performance of the proposed algorithm. We assume that the BS is located at $(0,0,0)$, the ARIS is placed at $(30,0,5)$, and the UE is located at $(70,10,0)$. The BS is equipped with $N=4$ FAs, the ARIS is equipped with $M=4$ reflecting elements, and the UE has a fixed single antenna. Additionally, we assume the communication frequency is 1.2 GHz, resulting in a wavelength $\lambda$ of 0.25 m.
We consider a geometric channel model where the transmission and reception paths are the same, i.e., $L_t^i=L_r^i=L=5, i\in\{\text{BR},\text{BU},\text{RU}\}$. The elevation and azimuth AoDs/AoAs are assumed to be i.i.d. variables with a uniform distribution, i.e., $\theta_{t,s_i}^i\sim\mathcal{U}[0,\pi], \varphi_{t,s_i}^i\sim\mathcal{U}[0,\pi], \theta_{r,k_i}^i\sim\mathcal{U}[0,\pi], \varphi_{r,k_i}^i\sim\mathcal{U}[0,\pi], i\in\{\text{BR},\text{BU},\text{RU}\}$. Furthermore, we consider the line-of-sight (LoS) links between the BS and the ARIS and between the ARIS and the UE, and model the corresponding path response matrices as $\bm{\Sigma}_{k}[1,1] \sim \mathcal{CN}(0,K_0 \left( \frac{d_k}{d_0}\right)^{-\alpha}\iota/(\iota+1)), k\in\{\text{BR}, \text{RU}\}$ and non line-of-sight (NLoS) link between BS and UE as $\bm{\Sigma}_{k}[b,b]\sim\mathcal{CN}(0,K_0 \left( \frac{d_k}{d_0}\right)^{-\alpha}/((\iota+1)(L-1))), b=2, 3, \cdots, L$, where $\iota=0.5$ denotes the ratio of the average power for LoS paths to that for NLoS paths \cite{HWu2024}. For the links between the BS and UE, we only consider the NLoS component, i.e., $\bm{\Sigma}_{\text{BU}}[b,b]\sim\mathcal{CN}(0,K_0 \left( \frac{d_{\text{BU}}}{d_0}\right)^{-\alpha}/L), b=1, 2, 3, \cdots, L$. The distance-dependent path loss model is given by $K_0 \left( \frac{d}{d_0}\right)^{-\alpha}$, where $K_0=-30$ dB is the average channel gain at the reference distance $d_0=1$m , and $\alpha$ is the path loss exponent. The path loss exponents for the BS-ARIS links, ARIS-UE links, and BS-UE links are 2.2, 3, and 3, respectively. Moreover, we set the maximum transmit power at the BS to $P_0=20$ dBm, the maximum transmit power at the ARIS to $P_1=10$ dBm, and the noise power at both the UE and the ARIS to $\sigma_{\text{R}}^2=\sigma_{\text{u}}^2=-70$ dBm. The minimum distance between two antennas is $D=\frac{\lambda}{2}$ m, and the movable range of the FAs is $S_t=\left[-{\frac{A}{2}},{\frac{A}{2}}  \right] \times \left[-{\frac{A}{2}},{\frac{A}{2}}  \right]$ m, with $A=4\lambda$. Internal iteration accuracy for the antenna position optimization subproblem $\epsilon_1=10^{-5}$, Algorithm 1 has a convergence accuracy of $\epsilon_2=10^{-5}$ and Algorithm 2 has a convergence accuracy of $\epsilon_3=10^{-4}$.

\begin{figure}[t]
\centering
        \includegraphics[width=0.375\textwidth]{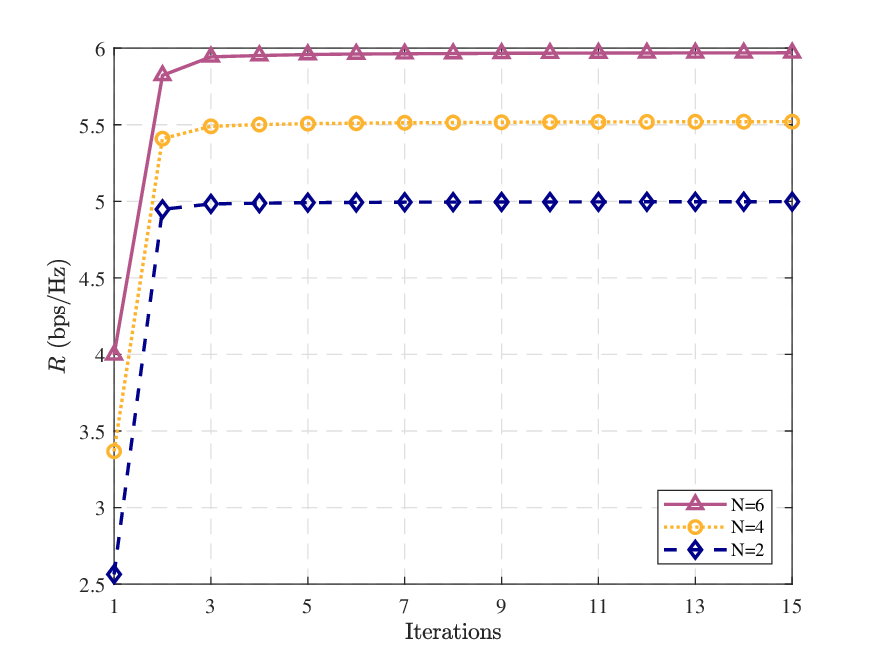}
        \caption{Convergence of the employed AO algorithm.}
        \label{fig1}
\end{figure}

In Fig.~\ref{fig1}, we present the convergence behavior of our employed AO algorithm, where the FAs' number of the BS $N=2$, $4$, and $6$. As we can see from Fig.~\ref{fig1}, our employed AO algorithm converges after roundly $4$ iterations for these three different settings, demonstrating its rapid convergence and validating the effectiveness of the algorithm. Furthermore, we observe that the achievable rate increases with the increasing number of FAs, which \textbf{\emph{effectively enhances the overall system performance and further validates the efficacy of FAS in communication systems.}}

\begin{figure}[t]
\centering
        \includegraphics[width=0.375\textwidth]{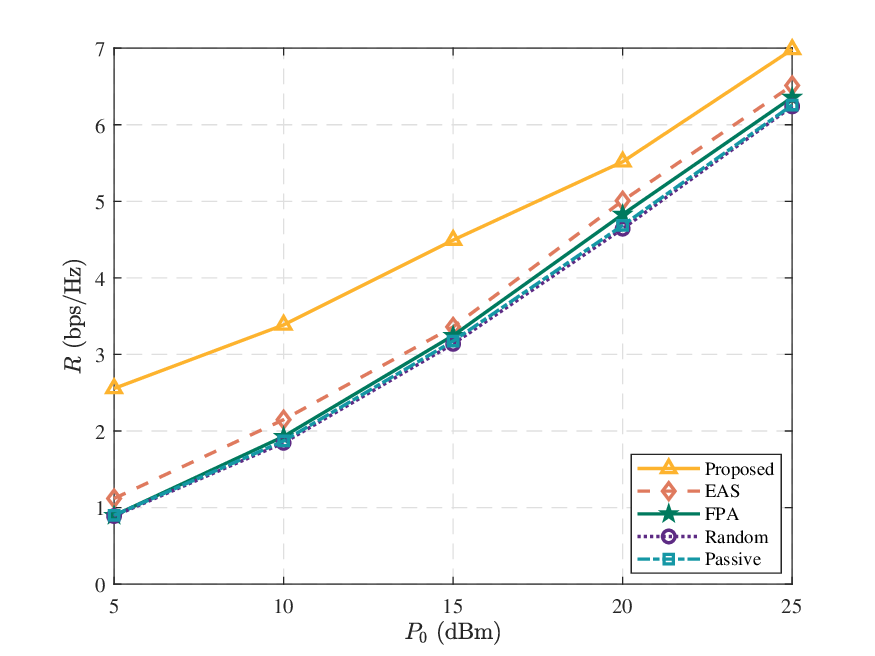}
        \caption{The maximum transmit power at the BS $P_{0}$ versus achievable rate $R$, where $N=4$. }
        \label{fig2}
\end{figure}

\begin{figure}[t]
\centering
        \includegraphics[width=0.375\textwidth]{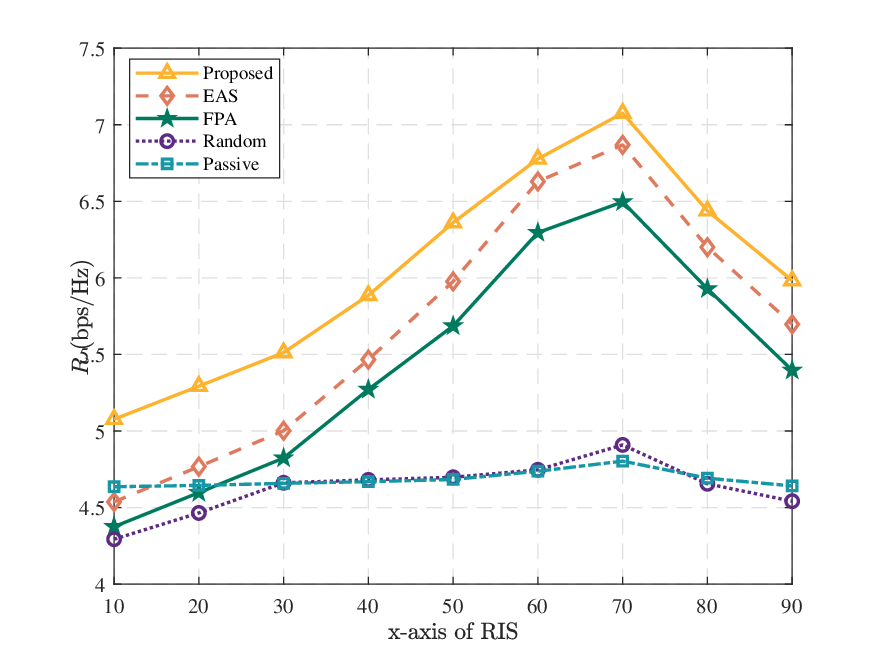}
        \caption{The x-axis of the ARIS versus achievable rate $R$, where $N=4$. }
        \label{fig4}
\end{figure}

\begin{figure}[t]
\centering
        \includegraphics[width=0.375\textwidth]{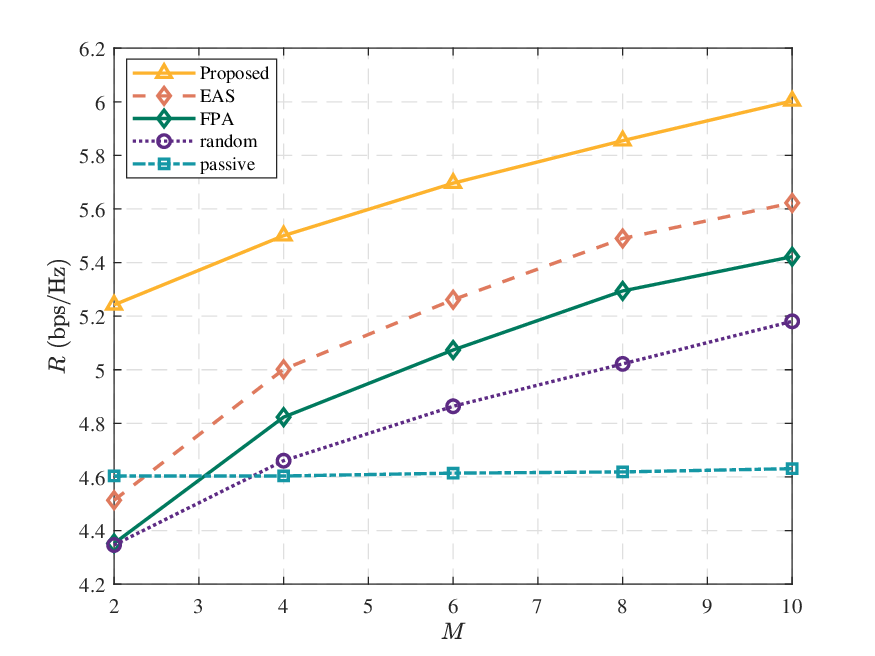}
        \caption{The number of reflecting elements within the  ARIS versus achievable rate $R$.}
        \label{fig5}
\end{figure}

\begin{figure}[t]
\centering
        \includegraphics[width=0.375\textwidth]{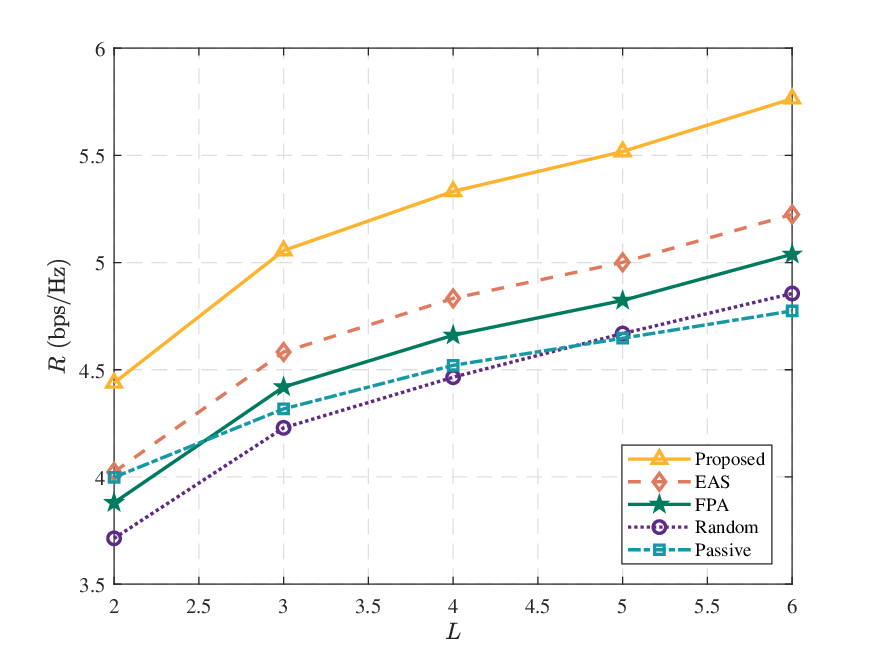}
        \caption{The number of transmission paths versus achievable rate $R$.}
        \label{fig6}
\end{figure}

\begin{figure}[t]
\centering
        \includegraphics[width=0.36\textwidth]{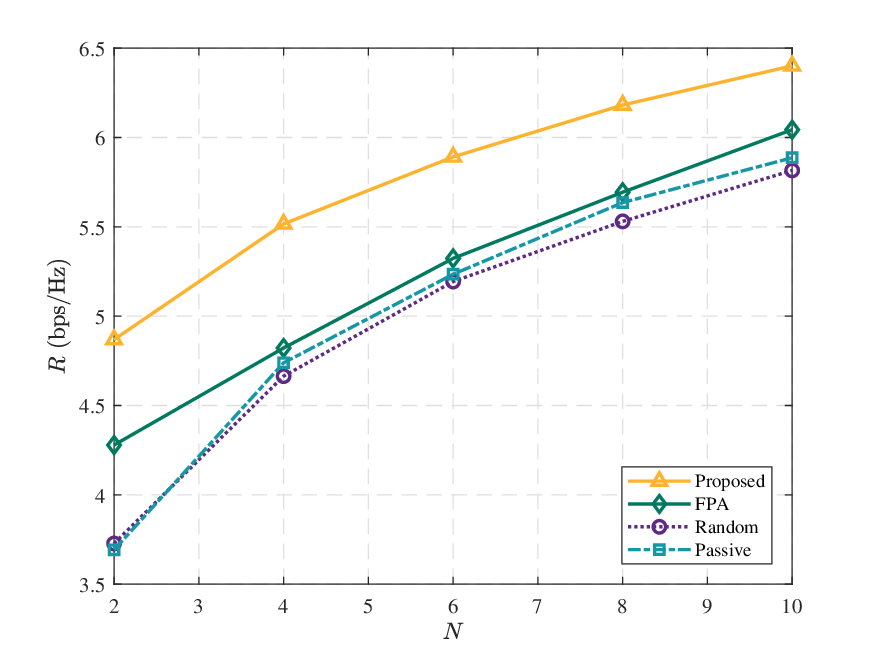}
        \caption{The number of antennas versus achievable rate $R$.}
        \label{fig7}
\end{figure}

\begin{figure}[t]
\centering
        \includegraphics[width=0.36\textwidth]{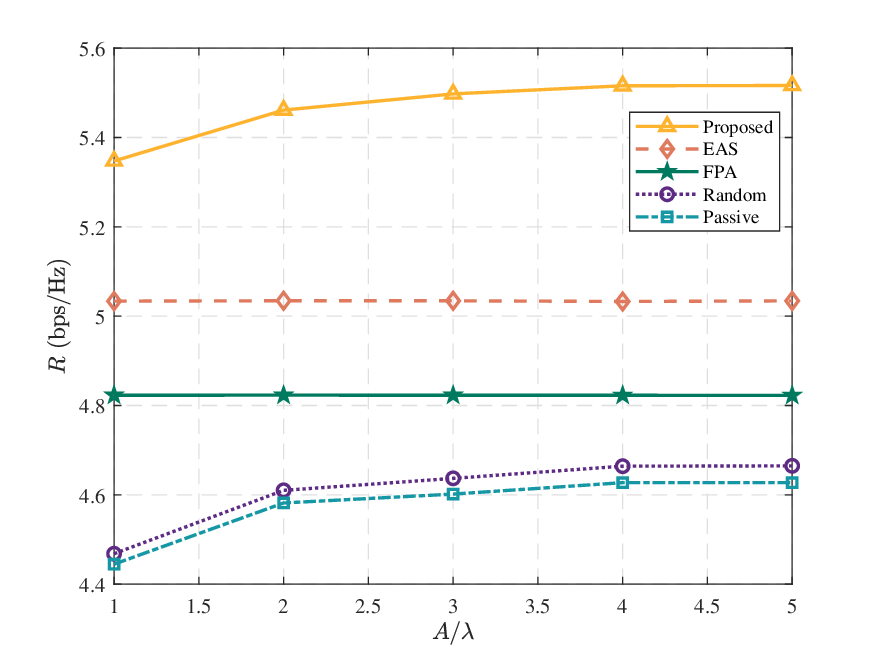}
        \caption{The normalized FAs' movable range $A/\lambda$ versus achievable rate $R$.}
        \label{fig8}
\end{figure}

In Fig.~\ref{fig2}, Fig.~\ref{fig4}, Fig.~\ref{fig5}, Fig.~\ref{fig6}, Fig.~\ref{fig7}, and Fig.~\ref{fig8}, we compare the proposed scheme with the following benchmarks:

$\mathbf{Exhaustive  \ antenna  \ selection   \ (EAS)}$: The BS is equipped with an FPA-based UPA, consisting of $2N$ antennas, from which $N$ antennas are selected through an exhaustive search.

$\mathbf{FPA}$: The BS is equipped with $N$ FPAs, while UE is equipped with a single FPA.

$\mathbf{Random}$: The phase shift matrix within the ARIS is set randomly.

$\mathbf{Passive}$: The passive RIS is employed for simulations. For fair comparison, the number of passive RIS's reflect elements can be obtained as \cite{ARIS1}
\begin{align}
M_{\text{passive}}=\frac{P_1+M(P_C+P_{DC})}{P_C},
\end{align}
where $P_C=-10$ dBm, and $P_{DC}=-5$ dBm.

In Fig.~\ref{fig2}, we compare the achievable rate $R$ of various schemes under different maximum transmit power at the BS $P_{0}$, where $N=4$. It is observed that as the $P_{0}$ increases, the achievable rate of all schemes increases. This effect becomes more pronounced as $P_{0}$ grows larger, demonstrating that higher $P_{0}$ results in a faster improvement in transmission rate. Additionally, our proposed scheme consistently outperforms the other schemes. Specifically, at $P_0 = 10$ dBm, the ``Proposed" scheme is approximately $50\%$ higher than the other schemes. Further, at $P_0 = 15$ dBm, the ``Proposed" scheme is approximately $33.3\%$ higher than the other schemes. \textbf{\emph{These calculations demonstrate that the ``Proposed" scheme not only achieves a higher achievable rate but does so with a significant margin, particularly at lower transmit power levels.}}

Moreover, this performance comparison in Fig.~\ref{fig2} confirms that the proposed scheme is superior to cases without FAs, with random phase shifts, and with passive RIS, thereby \textbf{\emph{validating the effectiveness of the proposed system}}. The significant performance gains observed in the proposed scheme underline its capability to enhance system efficiency and reliability, making it a promising solution for future wireless communication systems. \emph{\textbf{The optimization of the FAs' positions, the precise phase adjustment of the RIS, and the role of ARIS all work synergistically to create a well-rounded system.}} This holistic approach effectively enhances the performance and robustness of the system, contributing positively to advancements in wireless communication.

In Fig.~\ref{fig4}, we present the achievable rate $R$ as the x-coordinate of the ARIS is varied from $10$ to $90$, with the UE located at $(70,10,0)$. A clear trend can be observed from Fig.~\ref{fig4}: as the ARIS moves closer to the UE, the system performance improves significantly. This improvement highlights the importance of optimal ARIS positioning in enhancing communication efficiency. \emph{\textbf{The results suggest that positioning the ARIS closer to the UE provides a substantial performance boost, offering valuable insights for future wireless communication system deployments.}}

Besides, comparing the ``Proposed" scheme with the ``FPA" scheme in Fig.~\ref{fig4}, it is evident that the use of FAS significantly enhances the performance of RIS in reflecting signals. \textbf{\emph{The FAS effectively compensates for the path loss typically associated with RIS reflections, particularly when the ARIS is optimally positioned.}} As we can see from Fig.~\ref{fig4}, although the ``EAS" scheme, which involves random antenna selection, performs well, it still does not match the performance improvements brought by FAS. This comparison underscores \textbf{\emph{the substantial impact of FAS on enhancing future communication systems' performance}}.

Moreover, as shown in Fig.~\ref{fig4}, the comparison between the ``Proposed" scheme and the passive RIS clearly demonstrates that the introduction of ARIS results in a more significant performance improvement. This suggests that \textbf{\emph{ARIS is better suited to complement FAS than passive RIS, creating a more robust and comprehensive wireless communication system.}} Additionally, ARIS offers the benefits of greater energy efficiency and ease of deployment, making it a practical choice for enhancing communication performance. The findings indicate that even straightforward adjustments, such as optimizing the ARIS position, can lead to substantial performance gains.

In Fig.~\ref{fig5}, we present the achievable rate $R$ as the reflecting elements' number $M$  is varied from $2$ to $10$. As shown in Fig.~\ref{fig5}, FAS and ARIS each contribute to enhancing the performance of communication systems, but their relative importance varies depending on the system configuration and environment. Generally, ARIS tends to provide more significant gains, particularly as the number of RIS elements increases. For example, in scenarios with a larger number of RIS elements, ARIS can improve the achievable rate by approximately $25\% \thicksim 40\%$, often surpassing the $10\%\thicksim15\%$ improvements brought by FAS under similar conditions.

Upon further analysis of Fig.~\ref{fig5}, it becomes clear that FAS plays a crucial role in enhancing system performance, particularly when the number of RIS elements is low  (e.g., $M = 2$) or when the RIS performance begins to approach saturation. In such scenarios, where adding more reflecting elements yields diminishing returns, FAS can still provide a meaningful $10\%\thicksim15\%$ increase in the achievable rate. This makes FAS an invaluable component in systems where ARIS alone might not fully optimize performance.
Moreover, from Fig.~\ref{fig5}, it is evident that in FAS-RIS systems, the role of FAS is highly significant. \textbf{\emph{This strong influence of FAS causes the system to nearly reach its upper performance bound, even with increases in the number of reflecting elements in the RIS. }} This underscores the critical role that FAS plays in maximizing system performance.

\textbf{\emph{Remark 1}}:
From the perspective of optimizing the achievable rate,  \textbf{\emph{FAS plays a more crucial role when the number of RIS elements is low (e.g., \( M = 2 \)) or when the RIS performance begins to approach saturation}}. In these scenarios, FAS provides significant enhancements to system performance. Conversely, \textbf{\emph{ARIS becomes more important as the number of RIS elements increases, leading to a substantial boost in overall system performance.}} Their combined use results in a \textbf{\emph{win-win situation}}, where each technology complements the other, creating a more robust and efficient communication system.

Fig.~\ref{fig6} presents the achievable rate $R$  as a function of the transmission path $L$. As the value of $L$ increases, the achievable rate generally improves across all schemes, indicating that the system benefits from the increased number of transmission paths.  Delve deeper analysis on Fig.~\ref{fig6}, the proposed scheme consistently outperforms all other schemes across the entire range of $L$. The achievable rate improvement is more pronounced as $L$ increases, suggesting that the combination of FAS and ARIS effectively enhances the system's ability to utilize multiple transmission paths, thereby boosting performance. Besides, the ``random" and ``passive" schemes show less significant improvements as $L$ increases. Their performance is notably lower even compared to the FPA and EAS schemes. The relatively flat performance curve of the ``passive" scheme suggests that without the active optimization provided by ARIS or the dynamic antenna selection of FAS, the system's ability to benefit from multiple transmission paths is limited. The superior performance of the proposed scheme across all values of $L$ highlights the critical role of combining FAS with ARIS in the FAS-ARIS communication systems. This combination allows the system to better exploit multiple transmission paths, leading to a consistently higher achievable rate. The significant performance gap between the proposed scheme and the Passive RIS scheme underscores the importance of active reconfigurability and antenna optimization in leveraging the benefits of increased transmission paths.

Notably, when $L = 2$, corresponding to scenarios with fewer scattering paths, FAS provides a significant performance boost, emphasizing its critical role in such environments. In contrast, schemes like ``Random" and ``passive" show less improvement as $L$ increases, indicating that without the dynamic optimization provided by FAS or ARIS, the system's ability to capitalize on multiple transmission paths is limited. Overall, we can provide the following remark.

\textbf{\emph{Remark 2}}: In scenarios with limited scattering paths (e.g., $L = 2$), FAS plays a more critical role in boosting system performance, making it more important than ARIS in such contexts. Conversely, as the number of transmission paths increases, ARIS becomes more significant in enhancing system performance due to its ability to better leverage the additional paths. Therefore, while both FAS and ARIS are essential, \textbf{\emph{the relative importance of FAS increases in environments with fewer transmission }} paths.

Fig.~\ref{fig7} demonstrates that as the number of antennas $N$ increases, the achievable rate $R$ consistently improves across all schemes. The ``proposed" scheme outperforms the others throughout the range of $N$, with the performance gap widening as the number of antennas grows. \textbf{\emph{This clearly highlights the effectiveness of the proposed FAS and ARIS combination in maximizing system performance, particularly as more antennas are deployed. }}In contrast, the ``random" and ``passive" schemes exhibit limited improvement, emphasizing the critical role of active optimization strategies in fully capitalizing on the benefits of additional antennas. The FPA scheme also shows steady gains, but it still falls short of the performance achieved by the proposed method, especially at higher antenna counts.

In Fig.~\ref{fig8}, we investigate the impact of the normalized FAs' movable range $A/\lambda$ on the achievable rate $R$, with $M = 4$. The results indicate that as $A/\lambda$ increases, the achievable rates of all schemes, except for the ``FPA" and ``EAS" schemes, show significant improvement. The stagnation observed in the ``FPA" and ``EAS" schemes is likely due to the fixed positioning of the antennas, which limits their ability to benefit from a larger movable range. Notably, the achievable rates for all schemes except ``FPA" converge when $A/\lambda$ exceeds 3, suggesting that the system reaches its maximum performance within a limited movable range. This indicates that beyond a certain threshold, increasing the movable range does not yield further gains. Furthermore, the ``proposed" scheme consistently outperforms other benchmarks, demonstrating the superior performance achieved by effectively utilizing FAs in conjunction with ARIS.\textbf{\emph{ This insight underscores the importance of optimizing the movable range to fully harness the potential of FAS-ARIS systems.}}

\section{Conclusion}
This paper explored the question of which technology: FAS or  ARIS, plays a more crucial role in enhancing wireless communication systems. To address this, we developed a comprehensive system model and approached the problem from an optimization perspective by using MM, SCA, and SRCR algorithms. Our simulation results revealed that the relative importance of FAS and ARIS varies depending on the scenario: FAS is more critical in simpler models with fewer reflecting elements or limited transmission paths, while ARIS becomes more significant in complex scenarios with a higher number of reflecting elements or transmission paths. Besides, the integration of both FAS and ARIS has proved to be a win-win approach, resulting in a more robust and efficient communication system.

\appendices

\section{Proof of  \eqref{012-2}} \label{A1}
Since the rank-one constraint is relaxed when solving the problem in CVX, the solution $\mathbf{\widehat{W}}$ obtained from Problem \eqref{011} is usually not rank-one. Here, we reconstruct the rank-one solution using $\mathbf{\widehat{W}}$. We have
\begin{align}\label{037}
\mathbf{\widetilde{W}}=\mathbf{\widehat{W}}, \mathbf{\widetilde{W}}=\mathbf{\widetilde{w}}\mathbf{\widetilde{w}}^H, \mathbf{\widetilde{w}}=\left(\bm{\varpi} \mathbf{\widehat{W}} \bm{\varpi}^H      \right)^{-\frac{1}{2}}\mathbf{\widehat{W}}\bm{\varpi},
\end{align}
\begin{align}\label{038}
\bm{\varpi} \mathbf{\widetilde{W}} \bm{\varpi}^H=\bm{\varpi} \mathbf{\widetilde{w}}\mathbf{\widetilde{w}}^H \bm{\varpi}^H=\bm{\varpi} \mathbf{\widehat{W}} \bm{\varpi}^H.
\end{align}
Based on \eqref{037} and \eqref{038}, it is easy to prove that they satisfy constraints \eqref{011b} and \eqref{011c}. Furthermore, according to the Cauchy-Schwarz inequality \cite{HWu2024}, we have 
\begin{align}\label{039}
\left(\bm{\varpi} \mathbf{\widehat{W}} \bm{\varpi}^H \right)\left(  \mathbf{z} \mathbf{\widehat{W}} \mathbf{z}^H    \right)\geq |\mathbf{u} \mathbf{\widehat{W}} \bm{\varpi}^H|^2,
\end{align}
we can easily get
\begin{align}\label{040}
\mathbf{z} (\mathbf{\widehat{W}}-\mathbf{\widetilde{W}}) \mathbf{z}^H=\mathbf{z} \mathbf{\widehat{W}} \mathbf{z}^H-\left( \bm{\varpi} \mathbf{\widehat{W}} \bm{\varpi}^H   \right)^{-1}|\mathbf{z} \mathbf{\widehat{W}} \bm{\varpi}^H|^2\geq0,
\end{align}
which means
\begin{align}\label{041}
\mathbf{z} \mathbf{\widehat{W}} \mathbf{z}^H\geq\mathbf{z} \mathbf{\widetilde{W}} \mathbf{z}^H.
\end{align}
By comparing \eqref{041} and \eqref{037}, it is evident that the reconstructed rank-one solution necessarily satisfies all the constraints of Problem \eqref{011}.

\section{Derivations of   Hessian matrix} \label{A2}
Firstly, we define
\begin{align}
\nu^1_{s,k}(\mathbf{t}_n)&=\frac{2\pi}{\lambda}\left(\rho_{t,s}^\text{BU}(\mathbf{t}_n)- \rho_{t,k}^\text{BU}(\mathbf{t}_n)   \right)+\angle[\bm{\Omega}]_{s,k},\\
\nu^2_{s,k}(\mathbf{t}_n)&=\frac{2\pi}{\lambda}\left(\rho_{t,s}^\text{BR}(\mathbf{t}_n)- \rho_{t,k}^\text{BR}(\mathbf{t}_n)   \right)+\angle[\bm{\Xi}]_{s,k},\\
\nu^3_{s,k}(\mathbf{t}_n)&=\frac{2\pi}{\lambda}\left(\rho_{t,s}^\text{BR}(\mathbf{t}_n)- \rho_{t,k}^\text{BU}(\mathbf{t}_n)   \right)+\angle[\bm{\omega}]_{k}-\angle[\bm{\xi}]_{s},\\
\varsigma^1_{s}(\mathbf{t}_n)&=\frac{2\pi}{\lambda}\rho_{t,s}^\text{BR}(\mathbf{t}_n)-\angle[\bm{\xi}]_{s}+\angle\widetilde{\alpha},\\
\varsigma^2_{k}(\mathbf{t}_n)&=\frac{2\pi}{\lambda}\rho_{t,k}^\text{BU}(\mathbf{t}_n)-\angle[\bm{\omega}]_{k}+\angle\widetilde{\alpha},
\end{align}
where $\bm{\Omega}=\bm{\omega}\bm{\omega}^H,\bm{\Xi}=\bm{\xi}\bm{\xi}^H$.
According to Eq. \eqref{014}, we can provide the specific expression for $g(\mathbf{t}_n)$ as
\begin{align}
&g(\mathbf{t}_n)\nonumber\\
&=2\sum\limits_{s=1}^{L_t^\text{BU}-1}\sum\limits_{k=s+1}^{L_t^\text{BU}}\bigg\{[\mathbf{Q}]_{n,n}|[\bm{\Omega}]_{s,k}|\cos\left(\nu^1_{s,k}(\mathbf{t}_n)\right)\nonumber\\
&+\sum\limits_{s=1}^{L_t^\text{BU}}[\mathbf{Q}]_{n,n}|[\bm{\Omega}]_{s,s}|\nonumber\\
&+2\sum\limits_{s=1}^{L_t^\text{BR}-1}\sum\limits_{k=s+1}^{L_t^\text{BR}}[\mathbf{Q}]_{n,n}|[\bm{\Xi}]_{s,k}|\cos\left(\nu^2_{s,k}(\mathbf{t}_n)\right)\bigg\}\nonumber \\
&+\sum\limits_{s=1}^{L_t^\text{BU}}\bigg\{[\mathbf{Q}]_{n,n}|[\bm{\Xi}]_{s,s}|
+2\sum\limits_{s=1}^{L_t^\text{BR}}\sum\limits_{k=1}^{L_t^\text{BU}}[\mathbf{Q}]_{n,n}|[\bm{\xi}]_{s}| |[\bm{\omega}]_{k}|\cos\left(\nu^3_{s,k}(\mathbf{t}_n)\right)\nonumber\\
&+2\sum\limits_{s=1}^{L_t^\text{BR}}|\widetilde{\alpha}| |[\bm{\xi}]_s|\cos\left(\varsigma^1_{s}(\mathbf{t}_n)\right)+2\sum\limits_{k=1}^{L_t^\text{BU}}|\widetilde{\alpha}| |[\bm{\omega}]_k|\cos\left(\varsigma^2_{k}(\mathbf{t}_n)\right)\bigg\}, \label{00001}
\end{align}

To obtain the global lower bound of $ g(\mathbf{t}_n)$, we need to derive the gradient vector and the Hessian matrix of $g(\mathbf{t}_n)$, which can be represented as \cite{WMa23,LZhu23}
\begin{align}
\nabla g(\mathbf{t}_n)=&\left[\frac{\partial g(\mathbf{t}_n)}{\partial x^{t}_n},\frac{\partial g(\mathbf{t}_n)}{\partial y^{t}_n}     \right],\\
\nabla^2 g(\mathbf{t}_n)=&
\begin{bmatrix}
\frac{\partial g(\mathbf{t}_n)}{\partial x^{t}_n \partial x^{t}_n} & \frac{\partial g(\mathbf{t}_n)}{\partial x^{t}_n \partial y^{t}_n} \\
\frac{\partial g(\mathbf{t}_n)}{\partial y^{t}_n \partial x^{t}_n} & \frac{\partial g(\mathbf{t}_n)}{\partial y^{t}_n \partial y^{t}_n}
\end{bmatrix},
\end{align}
respectively. Among them, the mixed partial derivatives $\frac{\partial g(\mathbf{t}_n)}{\partial x^{t}_n \partial y^{t}_n}$ and $\frac{\partial g(\mathbf{t}_n)}{\partial y^{t}_n \partial x^{t}_n}$ are same.

Next, we will provide a specific expression. Details are presented as follows:

 \begin{align}
&\frac{\partial g(\mathbf{t}_n)}{\partial x^{t}_n}\nonumber\\
&=-\frac{4\pi}{\lambda}\sum\limits_{s=1}^{L_t^\text{BU}-1}\sum\limits_{k=s+1}^{L_t^\text{BU}}[\mathbf{Q}]_{n,n}|[\bm{\Omega}]_{s,k}|\nonumber\\
&\quad\times\left(-\cos\varphi^{\text{BU}}_{t,s}\sin\theta^{\text{BU}}_{t,s}+ \cos\varphi^{\text{BU}}_{t,k}\sin\theta^{\text{BU}}_{t,k}            \right)\sin\left( \nu^1_{s,k}(\mathbf{t}_n) \right)\nonumber \\
&\quad-\frac{4\pi}{\lambda}\sum\limits_{s=1}^{L_t^\text{BR}-1}\sum\limits_{k=s+1}^{L_t^\text{BR}}[\mathbf{Q}]_{n,n}|[\bm{\Xi}]_{s,k}|
\nonumber \\&\quad\times\left(-\cos\varphi^{\text{BR}}_{t,s}\sin\theta^{\text{BR}}_{t,s}+ \cos\varphi^{\text{BR}}_{t,k}\sin\theta^{\text{BR}}_{t,k}            \right)\sin\left( \nu^2_{s,k}(\mathbf{t}_n) \right)\nonumber \\
&\quad-\frac{4\pi}{\lambda}\sum\limits_{s=1}^{L_t^\text{BR}}\sum\limits_{k=1}^{L_t^\text{BU}}[\mathbf{Q}]_{n,n}|[\bm{\xi}]_{s}| |[\bm{\omega}]_{k}|
\nonumber \\&\quad\times\left(-\cos\varphi^{\text{BR}}_{t,s}\sin\theta^{\text{BR}}_{t,s}+ \cos\varphi^{\text{BU}}_{t,k}\sin\theta^{\text{BU}}_{t,k}            \right)\sin\left( \nu^3_{s,k}(\mathbf{t}_n) \right)\nonumber \\
&\quad-\frac{4\pi}{\lambda}\sum\limits_{s=1}^{L_t^\text{BR}}|\widetilde{\alpha}| |[\bm{\xi}]_s|\cos\varphi^{\text{BR}}_{t,s}\sin\theta^{\text{BR}}_{t,s}\sin\left(\varsigma^1_{s}(\mathbf{t}_n)\right)
\nonumber \\&\quad-\frac{4\pi}{\lambda}\sum\limits_{k=1}^{L_t^\text{BU}}|\widetilde{\alpha}| |[\bm{\omega}]_k|\cos\varphi^{\text{BU}}_{t,k}\sin\theta^{\text{BU}}_{t,k}\sin\left(\varsigma^2_{k}(\mathbf{t}_n)\right),\label{0001a} \end{align}

\begin{figure*}[t]
\begin{align}
\frac{\partial g(\mathbf{t}_n)}{\partial y^{t}_n}=&-\frac{4\pi}{\lambda}\sum\limits_{s=1}^{L_t^\text{BU}-1}\sum\limits_{k=s+1}^{L_t^\text{BU}}[\mathbf{Q}]_{n,n}|[\bm{\Omega}]_{s,k}|
\left(-\cos\theta^{\text{BU}}_{t,s}+ \cos\theta^{\text{BU}}_{t,k}            \right)\sin\left( \nu^1_{s,k}(\mathbf{t}_n) \right)\nonumber \\
&-\frac{4\pi}{\lambda}\sum\limits_{s=1}^{L_t^\text{BR}-1}\sum\limits_{k=s+1}^{L_t^\text{BR}}[\mathbf{Q}]_{n,n}|[\bm{\Xi}]_{s,k}|
\left(-\cos\theta^{\text{BR}}_{t,s}+ \cos\theta^{\text{BR}}_{t,k}            \right)\sin\left( \nu^2_{s,k}(\mathbf{t}_n) \right)\nonumber \\
&-\frac{4\pi}{\lambda}\sum\limits_{s=1}^{L_t^\text{BR}}\sum\limits_{k=1}^{L_t^\text{BU}}[\mathbf{Q}]_{n,n}|[\bm{\xi}]_{s}| |[\bm{\omega}]_{k}|
\left(-\cos\theta^{\text{BR}}_{t,s}+ \cos\theta^{\text{BU}}_{t,k}            \right)\sin\left( \nu^3_{s,k}(\mathbf{t}_n) \right)\nonumber \\
&-\frac{4\pi}{\lambda}\sum\limits_{s=1}^{L_t^\text{BR}}|\widetilde{\alpha}| |[\bm{\xi}]_s|\cos\theta^{\text{BR}}_{t,s}\sin\left(\varsigma^1_{s}(\mathbf{t}_n)\right)-\frac{4\pi}{\lambda}\sum\limits_{k=1}^{L_t^\text{BU}}|\widetilde{\alpha}| |[\bm{\omega}]_k|\cos\theta^{\text{BU}}_{t,k}\sin\left(\varsigma^2_{k}(\mathbf{t}_n)\right),\label{0001b}
\end{align}
\hrulefill
\end{figure*}
\begin{figure*}[t]
\begin{align}
\frac{\partial g(\mathbf{t}_n)}{\partial x^{t}_n \partial x^{t}_n}
&=-\frac{8\pi^2}{\lambda^2}\sum\limits_{s=1}^{L_t^\text{BU}-1}\sum\limits_{k=s+1}^{L_t^\text{BU}}[\mathbf{Q}]_{n,n}|[\bm{\Omega}]_{s,k}|
\left(-\cos\varphi_{t,s}^{\text{BU}}\sin\theta_{t,s}^{\text{BU}}+ \cos\varphi_{t,k}^{\text{BU}}\sin\theta_{t,k}^{\text{BU}}    \right)^2\cos(\nu^1_{s,k}(\mathbf{t}_n))\nonumber\\
&-\frac{8\pi^2}{\lambda}\sum\limits_{s=1}^{L_t^\text{BR}-1}\sum\limits_{k=s+1}^{L_t^\text{BR}}[\mathbf{Q}]_{n,n}|[\bm{\Xi}]_{s,k}|
\left( -\cos\varphi_{t,s}^{\text{BR}}\sin\theta_{t,s}^{\text{BR}}+ \cos\varphi_{t,k}^{\text{BR}}\sin\theta_{t,k}^{\text{BR}}     \right)^2\cos(\nu^2_{s,k}(\mathbf{t}_n))\nonumber\\
&-\frac{8\pi^2}{\lambda^2}\sum\limits_{s=1}^{L_t^\text{BR}}\sum\limits_{k=1}^{L_t^\text{BU}}[\mathbf{Q}]_{n,n}|[\bm{\xi}]_{s}| |\bm{\omega}_k|\left(-\cos\varphi_{t,s}^{\text{BU}}\sin\theta_{t,s}^{\text{BU}}+ \cos\varphi_{t,k}^{\text{BR}}\sin\theta_{t,k}^{\text{BR}}      \right)^2\cos(\nu^3_{s,k}(\mathbf{t}_n))\nonumber\\
&-\frac{8\pi^2}{\lambda^2}\sum\limits_{s=1}^{L_t^\text{BR}}|\widetilde{\alpha}| [\bm{\xi}]_{s}| \cos^2\varphi_{t,s}^{\text{BR}}\sin^2\theta_{t,s}^{\text{BR}}\cos(\varsigma^1_{s}(\mathbf{t}_n))-\frac{8\pi^2}{\lambda^2}\sum\limits_{k=1}^{L_t^\text{BU}}|\widetilde{\alpha}| [\bm{\omega}]_{k}|\cos^2\varphi_{t,k}^{\text{BU}}\sin^2\theta_{t,k}^{\text{BU}}\cos(\varsigma^2_{k}(\mathbf{t}_n)),\label{0002b}\\
\hrulefill
\frac{\partial g(\mathbf{t}_n)}{\partial y^{t}_n \partial y^{t}_n}
 =&-\frac{8\pi^2}{\lambda^2}\sum\limits_{s=1}^{L_t^\text{BU}-1}\sum\limits_{k=s+1}^{L_t^\text{BU}}[\mathbf{Q}]_{n,n}|[\bm{\Omega}]_{s,k}|
\left(-\cos\theta^{\text{BU}}_{t,s}+\cos\theta_{t,k}^{\text{BU}}      \right)^2\cos(\nu^1_{s,k}(\mathbf{t}_n))
\nonumber\\
&-\frac{8\pi^2}{\lambda^2}\sum\limits_{s=1}^{L_t^\text{BR}-1}\sum\limits_{k=s+1}^{L_t^\text{BR}}[\mathbf{Q}]_{n,n}|[\bm{\Xi}]_{s,k}|
\left(-\cos\theta_{t,s}^{\text{BR}}+\cos\theta_{t,k}^{\text{BR}}      \right)^2\cos(\nu^2_{s,k}(\mathbf{t}_n))
\nonumber\\&-\frac{8\pi^2}{\lambda^2}\sum\limits_{s=1}^{L_t^\text{BR}}\sum\limits_{k=1}^{L_t^\text{BU}}[\mathbf{Q}]_{n,n}|[\bm{\xi}]_{s}| |\bm{\omega}_k|\left(-\cos\theta_{t,s}^{\text{BU}}+\cos\theta_{t,k}^{\text{BR}}      \right)^2\cos(\nu^3_{s,k}(\mathbf{t}_n))\nonumber\\
&-\frac{8\pi^2}{\lambda^2}\sum\limits_{s=1}^{L_t^\text{BR}}|\widetilde{\alpha}| [\bm{\xi}]_{s}| \cos^2\theta_{t,s}^{\text{BR}}\cos(\varsigma^1_{s}(\mathbf{t}_n))-\frac{8\pi^2}{\lambda^2}\sum\limits_{k=1}^{L_t^\text{BU}}|\widetilde{\alpha}| [\bm{\omega}]_{k}| \cos^2\theta_{t,k}^{\text{BU}}\cos(\varsigma^2_{k}(\mathbf{t}_n)),\label{0002a}
\end{align}
\end{figure*}
\begin{figure*}[!t]
\begin{align}
\frac{\partial g(\mathbf{t}_n)}{\partial x^{t}_n \partial y^{t}_n}=&-\frac{8\pi^2}{\lambda^2}\sum\limits_{s=1}^{L_t^\text{BU}-1}\sum\limits_{k=s+1}^{L_t^\text{BU}}[\mathbf{Q}]_{n,n}|[\bm{\Omega}]_{s,k}|\left(-\cos\varphi_{t,s}^{\text{BU}}\sin\theta_{t,s}^{\text{BU}}+ \cos\varphi_{t,k}^{\text{BU}}\sin\theta_{t,k}^{\text{BU}}    \right)\left(-\cos\theta_{t,s}^{\text{BU}}+\cos\theta_{t,k}^{\text{BU}}       \right)\cos(\nu^1_{s,k}(\mathbf{t}_n))\nonumber \\
&-\frac{8\pi^2}{\lambda^2}\sum\limits_{s=1}^{L_t^\text{BR}-1}\sum\limits_{k=s+1}^{L_t^\text{BR}}[\mathbf{Q}]_{n,n}|[\bm{\Xi}]_{s,k}|\left( -\cos\varphi_{t,s}^{\text{BR}}\sin\theta_{t,s}^{\text{BR}}+ \cos\varphi_{t,k}^{\text{BR}}\sin\theta_{t,k}^{\text{BR}}     \right)\left(-\cos\theta_{t,s}^{\text{BR}}+\cos\theta_{t,k}^{\text{BR}}  \right)\cos(\nu^2_{s,k}(\mathbf{t}_n))\nonumber \\
&-\frac{8\pi^2}{\lambda^2}\sum\limits_{s=1}^{L_t^\text{BR}}\sum\limits_{k=1}^{L_t^\text{BU}}[\mathbf{Q}]_{n,n}|[\bm{\xi}]_{s}| |\bm{\omega}_k| \left(-\cos\varphi_{t,s}^{\text{BU}}\sin\theta_{t,s}^{\text{BU}}+ \cos\varphi_{t,k}^{\text{BR}}\sin\theta_{t,k}^{\text{BR}}      \right) \left(-\cos\theta_{t,s}^{\text{BU}}+\cos\theta_{t,k}^{\text{BR}}  \right)\cos(\nu^3_{s,k}(\mathbf{t}_n))\nonumber \\
&-\frac{8\pi^2}{\lambda^2}\sum\limits_{s=1}^{L_t^\text{BR}}|\widetilde{\alpha}| [\bm{\xi}]_{s}| \cos\varphi_{t,s}^{\text{BR}}\sin\theta_{t,s}^{\text{BR}}\cos\theta_{t,s}^{\text{BR}}\cos(\varsigma^1_{s}(\mathbf{t}_n))
-\frac{8\pi^2}{\lambda^2}\sum\limits_{k=1}^{L_t^\text{BU}}|\widetilde{\alpha}| [\bm{\omega}]_{k}| \cos\varphi_{t,k}^{\text{BU}}\sin\theta_{t,k}^{\text{BU}}\cos\theta_{t,k}^{\text{BU}}\cos(\varsigma^2_{k}(\mathbf{t}_n)).\label{0002c}
\end{align}
\hrulefill
\end{figure*}
\newpage
\newpage
\section{Derivation of $\kappa_n$} \label{A3}
After obtaining the Hessian matrix, we have
\begin{align}
||\nabla^2 g(\mathbf{t}_n)||_2^2 &\leq ||\nabla^2 g(\mathbf{t}_n)||_F^2 \nonumber \\
&=\left(\frac{\partial g(\mathbf{t}_n)}{\partial x^{t}_n \partial x^{t}_n}\right)^2+\left(\frac{\partial g(\mathbf{t}_n)}{\partial x^{t}_n \partial y^{t}_n}\right)^2 \nonumber \\
&=\left(\frac{\partial g(\mathbf{t}_n)}{\partial y^{t}_n \partial x^{t}_n}\right)^2+\left(\frac{\partial g(\mathbf{t}_n)}{\partial y^{t}_n \partial y^{t}_n}\right)^2.
\end{align}
Furthermore, due to $||\nabla^2 g(\mathbf{t}_n)||_2\mathbf{I}_2 \succeq \nabla^2 g(\mathbf{t}_n)$, we can choose $\kappa_n$ as
\begin{align}
\kappa_n=&\frac{64\pi^2}{\lambda^2}\Bigg(\sum\limits_{s=1}^{L_t^\text{BU}-1}\sum\limits_{k=s+1}^{L_t^\text{BU}}[\mathbf{Q}]_{n,n}|[\bm{\Omega}]_{s,k}| \nonumber \\
&+\sum\limits_{s=1}^{L_t^\text{BR}-1}\sum\limits_{k=s+1}^{L_t^\text{BR}}[\mathbf{Q}]_{n,n}|[\bm{\Xi}]_{s,k}| \nonumber \\
&+\sum\limits_{s=1}^{L_t^\text{BR}}\sum\limits_{k=1}^{L_t^\text{BU}}[\mathbf{Q}]_{n,n}|[\bm{\xi}]_{s}| |\bm{\omega}_k| \Bigg) \nonumber \\
&+\frac{16\pi^2}{\lambda^2}\left(\sum\limits_{s=1}^{L_t^\text{BR}}|\widetilde{\alpha}| [\bm{\xi}]_{s}| +\sum\limits_{k=1}^{L_t^\text{BU}}|\widetilde{\alpha}| [\bm{\omega}]_{k}|     \right),
\end{align}
which satisfying $\kappa_n \geq ||\nabla^2 g(\mathbf{t}_n)||_2^2$, and also satisfying $\kappa_n\mathbf{I}_2 \succeq \nabla^2 g(\mathbf{t}_n)$.

\bibliographystyle{IEEEtran}
				\bibliography{myre}

\end{document}